\documentclass[journal,10pt,twocolumn]{IEEEtran}

\usepackage[T1]{fontenc}
\usepackage[utf8]{inputenc}
\usepackage{cite}
\usepackage{amsmath,amssymb,amsfonts}
\usepackage{graphicx}
\usepackage{textcomp}
\usepackage{xcolor}
\usepackage{booktabs}
\usepackage{tabularx}
\usepackage{makecell}
\usepackage{tikz}
\usepackage{url}
\usepackage[hidelinks]{hyperref}
\usepackage{cleveref}

\usetikzlibrary{arrows.meta,positioning,fit,backgrounds,calc}

\tikzset{
  mflow/.style={-{Latex[length=1.5mm,width=1.1mm]}, draw=black!65, line width=0.45pt},
  mback/.style={mflow, densely dashed, draw=black!45},
  mstep/.style={rectangle, rounded corners=2.5pt, draw=black!55, fill=blue!6,
                align=left, font=\scriptsize, inner sep=3pt, line width=0.4pt},
  mgate/.style={mstep, fill=orange!14, draw=orange!70!black},
  mout/.style={mstep, fill=green!12, draw=green!50!black},
  mrisk/.style={mstep, fill=red!8, draw=red!55!black, font=\scriptsize\itshape},
  mlbl/.style={font=\scriptsize, inner sep=1.5pt, fill=white, text=black!70},
  mhdr/.style={font=\scriptsize\bfseries, text=black!75},
  mnote/.style={font=\tiny\itshape, text=black!60, align=left, inner sep=1.5pt},
}
\newcommand{\sn}[1]{\textbf{#1}~}

\newcolumntype{Y}{>{\raggedright\arraybackslash}X}
\newcolumntype{L}[1]{>{\raggedright\arraybackslash}p{#1}}

\begin{document}

\title{Research Methodologies for Cybersecurity in\\Enterprise Environments: A Narrative Review,\\Synthesis and Executable Guide}

\author{Tran Duc Le\\
Department of Mathematics, Statistics \& Computer Science\\
University of Wisconsin-Stout Polytechnic\\
\texttt{let@uwstout.edu}}

\maketitle

\begin{abstract}
Enterprise cybersecurity research draws on a wider range of methods than any single
community routinely teaches: a study may need a systematic review, a design-science
artifact, an interview study, a controlled detection experiment, an attack-graph model, a
testbed deployment, or a randomised field trial, and increasingly several of these at
once. Researchers entering the field therefore face a selection problem before they face a
technical one. This paper addresses that problem in two ways. First, it is a narrative
review and synthesis of the methodological practice visible in a corpus of 151 works whose
identifiers were all checked against a registry before inclusion, organising that practice
into eleven families and reporting, for each, what the family answers, how strong the
supporting evidence is, and where it fails. Second, it converts each family into an
executable protocol: ordered steps, the instruments and formulas the steps require, the
evaluation criteria that make a result defensible, the validity threats that most often
invalidate it, and a reporting checklist. Every protocol is also drawn, so that the
sequence of steps, the decisions that branch it and the threats that attach to each step
are legible before the prose is read. We also treat the contradictions in the
literature as evidence in their own right. The reported ranking of intrusion-detection
algorithms is inconsistent across studies that are individually careful: one comparison
places random
forest at 99.9\% accuracy above convolutional and recurrent networks at 98\%, another
places deep feed-forward networks first, and a third places recurrent networks
first. We argue that the pattern is most parsimoniously explained by evaluation design
rather than by the algorithms, because the studies differ in every design dimension
known to move the reported number by more than the margins that separate them. The evidence supports methodological pluralism disciplined by explicit
validity reasoning: match the design to the decision under study, triangulate technical
against organisational evidence, state the population a result generalises to, and report
the conditions under which the result would not hold.
\end{abstract}


\section{Introduction}
\label{sec:intro}

\IEEEPARstart{E}{nterprise} cybersecurity is the protection of an organisation's
information, systems, networks, applications, operational technology and cyber-physical
assets against events that would compromise confidentiality, integrity, availability,
safety or continuity of business. Modern enterprises depend on distributed information
systems, cloud services, container platforms and networked operational technology. Those
dependencies enlarge the attack surface and, more importantly for a researcher, create
relationships in which a weakness in one component determines the consequences in another.
Critical infrastructure makes the point sharply, because a cyber event there propagates
into physical process and operational disruption~\cite{jiang2023modelbased,koutsoukos2018sure}.

Research in this setting is correspondingly broad. It includes security assessment,
penetration testing, threat modeling, attack simulation, risk quantification, security
assurance, incident response, security operations, human factors, training and
governance. Assessment has been characterised as the activity that establishes assurance
that vital cyber assets are effectively protected, and threat modeling as the systematic
identification of vulnerabilities, threats, attack paths and
countermeasures~\cite{leszczyna2021review,wong2023security}. Cyber-threat intelligence
extends this by turning threat information into intelligence that supports organisational
decisions, although its uptake has remained concentrated in information-technology
operations rather than across enterprise
practice~\cite{ainslie2023cyberthreat}.

\subsection{The problem this paper addresses}

The methodological diversity of the field is not itself a problem. The problem is that the
families evaluate different outcomes and rarely explain themselves to one another.
Accuracy and $F_1$ dominate machine-learning work; incident-response studies measure
coordination, response speed and organisational learning; risk modeling reports
probabilities, resilience or time-to-compromise; training research measures engagement and
knowledge retention. A researcher who has been trained in one of these traditions has
little basis for judging when it is the wrong instrument. The consequence is visible in the
literature: studies that answer a question nobody asked, technically careful experiments
whose results do not transfer out of the benchmark that produced them, and organisational
studies whose conclusions cannot be checked.

The field has known this for a long time. An early analysis of why information-security
research was scarce reported that firm-level security is an unusually sensitive research
domain and documented a failed attempt at empirical validation as its principal
result~\cite{kotulic2004there}. Two decades later, a systematisation of knowledge on
security as a scientific pursuit argued that the field's difficulty is not a shortage of
technique but a shortage of clarity about what would count as
evidence~\cite{herley2017science}. This paper is written in that spirit and at a more
practical level.

\subsection{Scope}

We restrict attention to methodology, the design, conduct, evaluation and reporting of a
study, rather than to security techniques. A paper proposing a better detector is in
scope only insofar as it demonstrates how such proposals should be evaluated. We further
restrict attention to enterprise and organisational settings, including the
cyber-physical enterprises (grids, hospitals, transport, shipping, industrial process) that
serve as proxies where a general enterprise population is inaccessible.

We do not impose a publication-date window, nor do we characterise one. A
narrative review is selected for methodological standing rather than assembled to fill a
span of years. A date range describes the corpus without licensing statements
about the period, as a selective review cannot offer comprehensive coverage
between endpoints. Most of the works discussed here are
recent, the corpus median publication year is 2020 and 58 of the 151 works date from
2023 onward (\Cref{tab:corpus}), because most of the methodological argument is; the
foundational statements are
included wherever a current method rests on one, and are identified as such where they
appear.

\subsection{Contributions}

\begin{enumerate}
  \item A methodology taxonomy of eleven families for enterprise cybersecurity research,
        each mapped to the class of research question it answers and the class of claim it
        licenses (\Cref{sec:taxonomy}).
  \item An executable protocol for each family, ordered steps, instruments, formulas,
        evaluation criteria, validity threats, observed failure modes and a reporting
        checklist (\Cref{sec:secondary}--\Cref{sec:training}).
  \item A drawn form of each protocol, in a single visual vocabulary, showing the ordered
        steps, the decisions that branch them, the artifact each step must produce and the
        validity threat that attaches to it (\Cref{fig:selection}--\Cref{fig:recipes}). The
        figures are the part of this paper intended to be usable at the point of designing
        a study, rather than read once.
  \item A worked analysis of the clearest contradiction in the corpus, the inconsistent
        ranking of intrusion-detection algorithms, showing that the contradiction cannot
        be resolved by choosing a winner and that headline benchmark accuracy is
        therefore not a stable estimate of enterprise performance
        (\Cref{sec:contradiction}).
  \item Three end-to-end study recipes that combine families, for readers who want a
        starting template rather than a taxonomy (\Cref{sec:recipes}).
  \item A reference corpus in which every entry was verified against a registry
        before inclusion, with the procedure and its failures reported
        (\Cref{sec:verify-proc}).
\end{enumerate}

\subsection{How to read this paper}

A reader who already knows which method they need should go directly to that family's
section; the sections are self-contained, each carries a figure of its protocol, and each
ends with a checklist. The figures share one vocabulary throughout: a blue box is a step,
an amber box is a decision the protocol forces you to make explicitly, a green box is an
artifact the step must produce, and a red italic box is the validity threat that attaches
to that step. A dashed arrow is an iteration back to an earlier step.
A reader who does not should read \Cref{sec:taxonomy}, which contains the selection
procedure, and then follow it. A reader who is reviewing someone else's work will find
\Cref{sec:crosscutting} (validity, reproducibility, ethics, claim discipline) and the
checklists in Appendix~\ref{app:checklists} the most directly usable parts.

By \emph{executable} we mean followable: ordered, gated, and with each step's required
output stated, so that a study can be designed against them directly. The protocols are
not software; the checklists and figures are the form in which they are meant to be used.

\section{How This Review Was Conducted}
\label{sec:method}

Family~A (\S\ref{sec:secondary}) requires a review to report its own procedure;
accordingly, this section states the review type and why it was chosen, how the literature
was found and screened, how references were verified before inclusion, and how the
synthesis was produced.

\subsection{Review type, and why this one}
\label{sec:reviewtype}

This is a \emph{narrative review and synthesis}, and the label records a methodological
commitment. Snyder's account of literature review as a research
methodology in its own right distinguishes the systematic, semi-systematic and integrative
forms by the question each can answer: a systematic review answers a narrow question where
the primary studies are commensurable enough to be appraised against a common instrument,
whereas an integrative or narrative synthesis is the appropriate form when the purpose is
to map and critique a body of work that spans research traditions~\cite{snyder2019literature}.
Our question, which methodology should be used for which enterprise-security research
decision, and what does each license, is of the second kind. The primary studies here
include controlled detection experiments, interview studies, Delphi panels, simulations and
maturity models. There is no shared outcome measure across them and no risk-of-bias
instrument that applies to all of them, so a systematic protocol would either exclude most
of the field or apply an appraisal instrument outside its domain of validity.

The narrative form has its own standards, and they are not the absence of standards. Green
et al.'s guidance for narrative reviews in peer-reviewed journals fixes the obligations:
state the purpose, state how the literature was found, and make the basis of selection
visible to the reader rather than implicit in the
author's expertise~\cite{green2006writing}. Ferrari adds the requirement that a narrative
review declare its own susceptibility to selection according to the author's
viewpoint~\cite{ferrari2015writing}, which is the specific weakness of the form. Barry
et al.'s six-step approach to state-of-the-art review supplies the operational structure we
follow, delimit the field, search, screen, extract, synthesise, and state what the
synthesis does not
cover~\cite{barry2022state}. Where a narrative synthesis needs the retrieval discipline of
a systematic one without its appraisal apparatus, Turnbull et al.'s systematic--narrative
hybrid describes exactly the combination used here: a documented, reproducible search
feeding an interpretive synthesis~\cite{turnbull2023systematic}. The RAMESES publication
standards for meta-narrative review are the closest existing reporting standard to what
this paper does, and we follow their reporting logic, rationale, scope, search process,
selection, appraisal approach, synthesis process, even though the review is not itself a
meta-narrative one~\cite{wong2013rameses,wong2014development}.

Three consequences follow, and we state them here rather than only in the limitations.
This review does not claim completeness of coverage. It does not claim that the eleven
families partition the field without remainder. And its confidence labels
(\Cref{sec:synthesis}) are reasoned judgements against stated definitions, not scores from
a validated instrument.

\subsection{Sources and retrieval}

We searched three bibliographic databases: OpenAlex (for
cross-publisher coverage and citation context), Semantic Scholar (for
semantically-related work), and arXiv (for methodological work in security
that appears as a preprint before, or instead of, a venue). General web search was not
used as a literature source. We queried one methodology facet at a time rather than as a
single broad query, for example \emph{cyber range testbed security experiments},
\emph{Delphi study cybersecurity experts consensus}, \emph{temporal bias malware
classification}, because long multi-clause queries returned unrelated work.

What this review does not report, and should: the complete search record, the wording of
each query, the date it was run and how many records it returned in each of the two
rounds, together with how many candidates were screened down to the 151 included works,
is not printed here and not yet released alongside the preprint. Family~A's protocol
(\Cref{sec:secondary}) requires a review to record and report exactly that, and this
version does not meet its own standard; the three example queries above illustrate the
style of search only. The repair is to release the full search record as supplementary
material; of the selection record, only the aggregate verification outcomes are
reported (\Cref{sec:verify-proc}), together with the two identifier failures the check
caught and their resolution.

A prior thematic synthesis prepared for this project contributed a seed set of 49 works
covering assessment, modeling, detection, human factors and operations. That seed set was
not accepted on trust: every one of its references was independently verified by the
procedure below, and its factual claims were checked against publisher-hosted abstracts
before any of them were carried into this paper. Section~\ref{sec:seed-audit} reports the
outcome of that audit, including the one part of the antecedent synthesis we declined to
reproduce.

\subsection{Inclusion criteria}

A work was included if it satisfied all of:
\begin{enumerate}
  \item \textbf{Security focus.} It addresses cybersecurity, information security or a
        directly adjacent security research problem, or it is a methodological
        foundation (for example inter-rater reliability, mapping-study guidelines) that
        security research demonstrably depends on.
  \item \textbf{Enterprise relevance.} It investigates an organisational, business,
        production-scale or cyber-physical enterprise setting, or supplies a method
        explicitly applicable to one.
  \item \textbf{Methodological explicitness.} It describes a research design, evaluation
        procedure or reproducible methodological approach, rather than presenting only an
        artifact or a threat survey.
  \item \textbf{Sufficient detail.} It reports enough about procedure, data, instruments
        or implementation to inform the design of a new study.
  \item \textbf{Stated evaluation.} It evaluates its own contribution against stated
        metrics, comparisons, case evidence or expert judgement.
\end{enumerate}

\subsection{Verification procedure}
\label{sec:verify-proc}

Before we included any reference, we checked it against its source record:

\begin{enumerate}
  \item For each Digital Object Identifier (DOI) we looked up the record and compared the first-author family name,
        publication year, title and container title against the citation we intended to
        record.
  \item We did the same for each arXiv identifier.
  \item We then built the bibliography from the source record itself, so that a metadata
        field could not silently diverge from the source.
  \item Numeric claims carried over from the seed synthesis were checked against the
        publisher-hosted abstract of the cited work.
\end{enumerate}

The check is inexpensive and it caught real errors. Two candidate identifiers failed the
check: one did not resolve at all, and one turned out to refer to an entirely different
paper than the one intended, a class of error that no amount of proofreading detects,
because the citation string looks correct. Both were excluded rather than repaired from
memory.

We searched and selected in two rounds, the second after the taxonomy was complete, which
is the point at which the gaps in the first round become visible: the reporting standards
each family should be written against, the methodology literature for the review form
itself (\Cref{sec:reviewtype}), and the primary-method exemplars for the families that the
first round had covered thinnest. All 41 works added in the second round checked out. The
second round also recovered one of the two earlier failures. The identifier that had
turned out to refer to an unrelated paper belonged to a work we still wanted, a
systematisation of benchmarking flaws in systems
security~\cite{vanderkouwe2019benchmarking}, and searching for the work rather than
repairing the string located the correct record. That is the right order in which to do
it, and the wrong order is what produces a citation that looks clean and points somewhere
else. Of the 151 works we included, 129 were verified against their DOI record and 22
against their arXiv record.

Five references carry a year discrepancy of one between the online-first date frequently
cited in secondary literature and the issue date of record (for example, a cyber-range
review cited as 2019 that carries a 2020 issue date). We use the issue date reported in
the registry throughout, which is why some years here differ by one from the same work
cited elsewhere.

\Cref{tab:corpus} records what can be counted directly from the verified bibliography:
the corpus spans 1977--2026 with a median publication year of 2020, and 58 of the 151
works date from 2023 onward. The distribution describes this review's selection, not the
field: the corpus was assembled by the judgement and the facet-by-facet search described
above, so these counts are provenance for the frequency characterisations in the family
sections, not estimates of the literature's composition.

\begin{table}[t]
\caption{Composition of the 151-work corpus, counted from the verified bibliography.}
\label{tab:corpus}
\centering
\footnotesize
\begin{tabular}{@{}lr@{}}
\hline
\textbf{Publication period} & \textbf{Works} \\
\hline
1977--2009 & 19 \\
2010--2014 & 16 \\
2015--2019 & 34 \\
2020--2022 & 24 \\
2023--2026 & 58 \\
\hline
\textbf{Venue type} & \textbf{Works} \\
\hline
Journal article & 97 \\
Conference paper & 25 \\
Preprint or other unindexed & 23 \\
Technical report & 2 \\
Book & 2 \\
Book chapter & 2 \\
\hline
\end{tabular}
\end{table}

\subsection{Synthesis}
\label{sec:synthesis}

For each work we recorded the methodology family, research design, enterprise context,
data sources, procedure, evaluation instrument, principal finding and stated limitations.
Synthesis was thematic. We did not compute a pooled effect estimate, because the outcome
measures across families, accuracy, time-to-compromise, maturity level, $\kappa$,
perceived usefulness, are not commensurable, and a single number would have had no
interpretation.

Each thematic claim in \Cref{sec:secondary}--\Cref{sec:training} carries a confidence
label. We use three levels, defined as follows.
\emph{Strong}: multiple independent studies, consistent direction, method transparent
enough to be checked.
\emph{Moderate}: consistent direction but with dependence on context, expert input or
single-site evidence.
\emph{Limited}: supported by one study, by formative evaluation only, or by evidence whose
population differs materially from the enterprise population the claim addresses.

The scheme is modelled on Grading of Recommendations, Assessment, Development and Evaluation (GRADE), which rates a body of evidence rather than an individual
study and separates the quality of that evidence from the strength of the recommendation
drawn from it~\cite{guyatt2008grade}. We borrow the separation and the practice of
recording the reason for each rating, but not the instrument: GRADE's downgrading criteria
assume comparable outcome measures across studies, which is precisely the condition this
corpus does not satisfy. The labels here should therefore be read as structured editorial
judgement, checkable against the definitions above, and not as instrument scores.

\subsection{Limitations of this review}
\label{sec:review-limits}

This is a narrative synthesis, not a meta-analysis, and it inherits six limitations.
First, and definitionally, the corpus is selected rather than exhaustive. A narrative
review is assembled by judgement, and judgement is where the author's own background enters
the result~\cite{ferrari2015writing}; ours is in security engineering and empirical
software engineering, which is visible in how much space the detection and experimentation
families receive relative to, say, security economics. A reader should treat the eleven
families as a defensible partition of what we found, not as a census of what exists.
Second, retrieval was facet-by-facet and relied on database search; a facet we did not
formulate is a facet we did not cover, and we name the ones we know we under-covered in
\Cref{sec:gaps}. Third, no formal risk-of-bias instrument was applied to the included
studies; the confidence labels are our reasoned judgement against the definitions above,
not an instrument score. Fourth, the corpus over-represents work that publishes its method,
which is a form of selection: a widely-used practice that nobody writes up is invisible
here. Fifth, we did not obtain the full text of every included work; where extraction
rested on an abstract or a structured record, the corresponding claim is labelled at most
\emph{Moderate}. Sixth, extraction and family assignment were performed by a single
coder. The frequency characterisations the family sections rely on (``mostly'',
``predominantly'') are one systematic reading of the 151 works, not counted or
reliability-tested estimates; \Cref{tab:corpus} reports what could be counted
directly from the verified bibliography, and a double-coded sample of the family
assignments is the obvious strengthening.

Because the first of these is the form's characteristic weakness rather than an incidental
one, we make the compensating move that the narrative-review guidance
recommends~\cite{green2006writing,barry2022state}: the retrieval procedure, the inclusion
criteria, the verification record and the explicit statement of what we under-covered are
all reported above and in \Cref{sec:gaps}, so that the selection is inspectable even though
it is not exhaustive.

\subsection{Audit of the antecedent synthesis}
\label{sec:seed-audit}

Because the seed synthesis materially shaped \Cref{sec:secondary}--\Cref{sec:training}, we
report its audit explicitly.

All 49 of its references checked out, and in every case the first author, title and venue
matched the citation. None was fabricated. Its load-bearing numeric claims also held
verbatim against the publishers' own abstracts: the comparison reporting 98\% accuracy for
convolutional and recurrent models against 99.9\% for random forest~\cite{ali2025deep}; the
hybrid ensemble reporting accuracy, precision, recall and $F_1$ approaching 100\% under
Synthetic Minority Over-sampling Technique (SMOTE), weighted soft voting and five-fold cross-validation, with independent benchmark
evaluation~\cite{almuhanna2025deep}; the awareness-metric framework derived from a review
of 32 papers adapting four indicators, impact, sustainability, accessibility,
monitoring~\cite{chaudhary2022developing}; and the four-week experience-sampling design
behind the within-person treatment of neutralisation theory~\cite{cram2024time}.

One part we declined to reproduce. The antecedent synthesis reported a study-selection
flow of 400 records identified, 200 screened, 150 excluded and 50 included, which implies
that no work was excluded at full-text assessment, an outcome that essentially does not
occur in a real screening process. Its own limitations section conceded that the stated
totals exceed the records whose data are presented. We therefore treat those figures as
unsupported and do not restate them, here or as a description of our own retrieval. The
episode is a useful illustration of the paper's own argument: a reference list can be
completely clean while the process narrative around it is not, so the two need separate
checks.

\section{A Taxonomy and a Selection Procedure}
\label{sec:taxonomy}

\subsection{Eleven families}

\Cref{tab:taxonomy} maps the families to the question each answers and the claim each
licenses. The final column is the one most often ignored: a method licenses a specific
form of statement, and a study that draws a stronger form has made an error that no amount
of technical care repairs.

\begin{table*}[t]
\caption{Methodology families, the questions they answer, and the claims they license.}
\label{tab:taxonomy}
\centering
\footnotesize
\begin{tabularx}{\textwidth}{@{}l Y Y Y@{}}
\toprule
\textbf{Family} & \textbf{Answers the question} & \textbf{Primary output} & \textbf{Licenses a claim of the form} \\
\midrule
Secondary research (\S\ref{sec:secondary}) & What is known, and where are the gaps? & Taxonomy, map, gap agenda & ``The published literature reports \ldots'' \\
Design science (\S\ref{sec:dsr}) & Can an artifact be built that solves this class of problem? & Artifact plus evaluation & ``This artifact achieved \ldots\ in this evaluation setting'' \\
Qualitative inquiry (\S\ref{sec:qualitative}) & Why do practitioners act as they do? & Themes, models, mechanisms & ``In these settings, participants described \ldots'' \\
Survey and behavioural (\S\ref{sec:survey}) & How is a construct distributed, and what predicts it? & Estimates, path coefficients & ``In this sampled population, X was associated with Y'' \\
Threat modeling and risk (\S\ref{sec:threat}) & Which paths exist, and what would they cost? & Threat set, attack graph, probability & ``Under this model and these inputs, \ldots'' \\
Simulation and model-based (\S\ref{sec:simulation}) & How does the system behave under attack? & Trajectories, sensitivities & ``In the modelled system, \ldots'' \\
Machine-learning experiment (\S\ref{sec:ml}) & Does this detector work, and under what conditions? & Metrics with confidence bounds & ``On this data, under this split, \ldots'' \\
Language-model and agentic evaluation (\S\ref{sec:llm}) & Is this language-model-based system fit for the task? & Benchmark result plus validity analysis & ``Under this uncontaminated protocol, \ldots'' \\
Measurement and telemetry (\S\ref{sec:measurement}) & What can actually be observed in practice? & Coverage and efficacy measures & ``With this telemetry, this fraction was detectable'' \\
Experimental infrastructure (\S\ref{sec:infrastructure}) & Where can this be tested safely and repeatably? & Testbed, range, emulation plan & ``This result is reproducible on this configuration'' \\
Field experiment and training (\S\ref{sec:training}) & Does the intervention change behaviour? & Effect estimate, learning outcome & ``Assignment to the intervention changed \ldots'' \\
\bottomrule
\end{tabularx}
\end{table*}

\subsection{Selection procedure}

Method follows the decision under study, not the researcher's training. The following
procedure is deliberately mechanical.

\begin{enumerate}
  \item \textbf{Write the claim you intend to end up making}, in one sentence, in the
        past tense, with its population and its conditions attached. If the sentence
        cannot be written, the study is not yet defined and no methodology can be chosen.
  \item \textbf{Classify the claim.} Descriptive (what exists), associational (what varies
        together), causal (what changes what), constructive (what can be built), or
        comparative (which of these is better under stated conditions).
  \item \textbf{Read the licensing column} of \Cref{tab:taxonomy} and eliminate every
        family whose licensed form is weaker than the claim you wrote.
  \item \textbf{Check feasibility of access.} Enterprise research is access-bound far more
        than it is technique-bound. Determine what you can actually obtain: production
        telemetry, practitioners' time, a testbed, a workforce to randomise, or only public
        artifacts. Eliminate families whose data you cannot get.
  \item \textbf{If more than one family survives, take two.} Triangulation across a
        technical and an organisational family is the recurring recommendation of this
        corpus, and it is what distinguishes the studies whose conclusions survive contact
        with practice.
  \item \textbf{Pre-specify the negative result.} State now what observation would make
        you report that the approach did not work. A design that has no such observation
        is not an evaluation.
\end{enumerate}

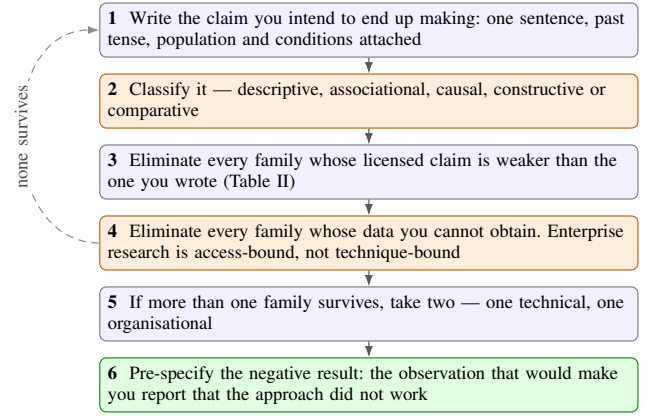
\begin{figure}[t]
\centering
\begin{tikzpicture}[
  node distance=2.1mm,
  mstep/.append style={text width=0.78\columnwidth},
  mgate/.append style={text width=0.78\columnwidth},
  mout/.append style={text width=0.78\columnwidth}]
\node[mstep] (s1) {\sn{1} Write the claim you intend to end up making: one sentence, past tense, population and conditions attached};
\node[mgate, below=of s1] (s2) {\sn{2} Classify it --- descriptive, associational, causal, constructive or comparative};
\node[mstep, below=of s2] (s3) {\sn{3} Eliminate every family whose licensed claim is weaker than the one you wrote (\Cref{tab:taxonomy})};
\node[mgate, below=of s3] (s4) {\sn{4} Eliminate every family whose data you cannot obtain. Enterprise research is access-bound, not technique-bound};
\node[mstep, below=of s4] (s5) {\sn{5} If more than one family survives, take two --- one technical, one organisational};
\node[mout, below=of s5] (s6) {\sn{6} Pre-specify the negative result: the observation that would make you report that the approach did not work};
\foreach \a/\b in {s1/s2,s2/s3,s3/s4,s4/s5,s5/s6} \draw[mflow] (\a) -- (\b);
\draw[mback] (s4.west) to[out=180,in=180,looseness=1.25]
  node[mlbl,pos=0.5,rotate=90]{none survives} (s1.west);
\end{tikzpicture}
\caption{The selection procedure of \S\ref{sec:taxonomy}, which runs before any method is
chosen. The two amber steps are the ones researchers skip: a claim that has not been
classified cannot be matched to a family, and a family whose data cannot be obtained is not
a candidate however well it fits. If step~4 eliminates everything, the claim, not the
method, is what has to change.}
\label{fig:selection}
\end{figure}

Step 6 deserves emphasis. Across the corpus, in our coding of it
(\Cref{sec:review-limits}), the studies whose findings we labelled
\emph{Limited} are mostly those in which the evaluation could only have come out
one way. \Cref{fig:selection} draws the procedure, including the loop that most study
designs need at least once: if no family survives the access check, the claim is the thing
that has to be renegotiated.

\section{Family A: Secondary Research}
\label{sec:secondary}

\subsection{Purpose and when to use it}

Secondary research synthesises existing studies. Use it when the question is about the
state of knowledge, what methods exist, how they are evaluated, what is missing, rather
than about the world directly. In this corpus it is the principal landscape-building
method, applied to assessment methods, machine-learning detection, cyber ranges,
penetration testing, threat intelligence, awareness programmes and organisational
implementation factors~\cite{leszczyna2021review,chinnasamy2025deep,yamin2020cyber,alhamed2023systematic,khaw2024building}.

Three variants are worth distinguishing. A \emph{systematic literature review} answers a
focused question and appraises the evidence. A \emph{systematic mapping study} answers a
broader question by categorising a field without deep appraisal, and is the right choice
when the field is too immature for synthesis~\cite{petersen2015guidelines}. A
\emph{systematisation of knowledge} (SoK), the security community's own form, reorganises
a field around a conceptual contribution rather than around a protocol; the SoK on
evaluations in industrial intrusion detection and the SoK on algorithmic red teaming are
recent instances~\cite{lamberts2023evaluations,srivastava2026systematic}.

\subsection{Protocol}

\begin{enumerate}
  \item \textbf{Formulate the question} in PICO-like terms adapted to computing:
        population (which systems, which organisations), intervention (which method or
        technique), comparison, outcome (which measured property).
  \item \textbf{Write and register the protocol before searching}, including the query
        strings, the databases, the inclusion and exclusion criteria, and the extraction
        form. Deviations discovered later are reported as deviations rather than
        retrofitted.
  \item \textbf{Search.} Use at least two indexing services with different coverage
        models. Record the exact query string, the service, the date and the hit count for
        each. In security, add the preprint servers explicitly: a substantial part of
        current methodological work appears there first.
  \item \textbf{Snowball.} Backward snowballing follows the reference lists of included
        works; forward snowballing follows their citers. Wohlin's guidelines make the
        start set and the iteration explicit and are the standard reference for doing this
        reproducibly~\cite{wohlin2014guidelines}. Snowballing routinely recovers work that
        keyword search misses because the community uses a different vocabulary.
  \item \textbf{Screen} in two stages, title/abstract, then full text, with a second
        screener on a sample, and report the agreement (\Cref{sec:kappa}). Record the
        exclusion reason at full-text stage for every excluded work.
  \item \textbf{Extract} using the pre-specified form. Pilot the form on five papers and
        revise before the full pass.
  \item \textbf{Synthesise.} For a review, thematic synthesis with an auditable coding
        step is the workable default; Cruzes and Dyb\aa{}'s recommended steps adapt it to
        engineering literature~\cite{cruzes2011recommended}. For a mapping study, produce
        the classification and the frequency distributions.
  \item \textbf{Report} against the Preferred Reporting Items for Systematic Reviews and Meta-Analyses (PRISMA) 2020 statement~\cite{page2021prisma}, including the flow of
        records through identification, screening, eligibility and inclusion.
\end{enumerate}

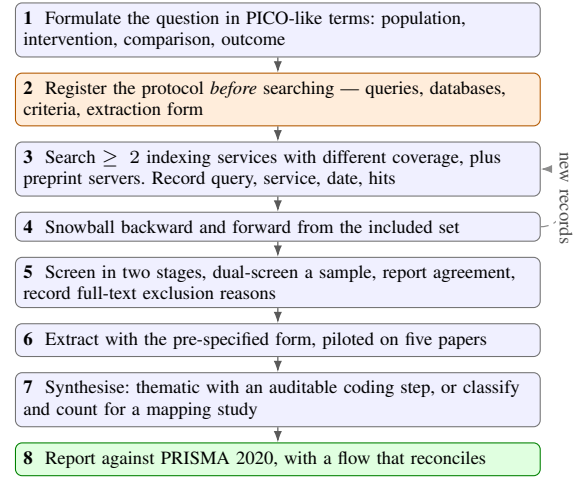
\begin{figure}[t]
\centering
\begin{tikzpicture}[
  node distance=2.0mm,
  mstep/.append style={text width=0.76\columnwidth},
  mgate/.append style={text width=0.76\columnwidth},
  mout/.append style={text width=0.76\columnwidth}]
\node[mstep] (a1) {\sn{1} Formulate the question in PICO-like terms: population, intervention, comparison, outcome};
\node[mgate, below=of a1] (a2) {\sn{2} Register the protocol \emph{before} searching --- queries, databases, criteria, extraction form};
\node[mstep, below=of a2] (a3) {\sn{3} Search $\geq 2$ indexing services with different coverage, plus preprint servers. Record query, service, date, hits};
\node[mstep, below=of a3] (a4) {\sn{4} Snowball backward and forward from the included set};
\node[mstep, below=of a4] (a5) {\sn{5} Screen in two stages, dual-screen a sample, report agreement, record full-text exclusion reasons};
\node[mstep, below=of a5] (a6) {\sn{6} Extract with the pre-specified form, piloted on five papers};
\node[mstep, below=of a6] (a7) {\sn{7} Synthesise: thematic with an auditable coding step, or classify and count for a mapping study};
\node[mout, below=of a7] (a8) {\sn{8} Report against PRISMA 2020, with a flow that reconciles};
\foreach \a/\b in {a1/a2,a2/a3,a3/a4,a4/a5,a5/a6,a6/a7,a7/a8} \draw[mflow] (\a) -- (\b);
\draw[mback] (a4.east) to[out=0,in=0,looseness=1.5]
  node[mlbl,pos=0.5,rotate=270]{new records} (a3.east);
\end{tikzpicture}
\caption{Family A. The two steps that decide whether the review is checkable are~2 and~5:
a protocol written after the search cannot constrain it, and a screening stage with no
recorded exclusion reasons produces a flow diagram that no reader can audit. Snowballing
feeds records back into screening, which is why the search cannot be closed before it.}
\label{fig:secondary}
\end{figure}

The protocol is illustrated in \Cref{fig:secondary}.

\subsection{Instruments}

The reporting standards worth knowing are PRISMA 2020 for reviews of any
kind~\cite{page2021prisma}, Kitchenham's software-engineering adaptation of systematic
review procedure~\cite{kitchenham2009systematic}, Petersen's mapping-study
guidelines~\cite{petersen2015guidelines}, and ROSES where an environmental-evidence style
protocol is preferred, as in the thematic cybersecurity framework
of~\cite{khaw2024building}.

Two companions to PRISMA are worth naming because they cover the cases security reviews
most often fall into. PRISMA-P specifies what belongs in the \emph{protocol}, the document
written at step~2, before any searching, and is the instrument that makes step~2 an
artifact rather than an intention~\cite{moher2015preferred}. PRISMA-ScR extends the
reporting items to scoping reviews, which is the correct label for a large share of what
the security literature calls a survey: a review that maps the extent and nature of a
field without appraising the strength of its evidence~\cite{tricco2018prisma}. Choosing
the standard is part of choosing the design, not a formatting decision taken at
submission.

\subsection{What the evidence supports}

\textbf{Confidence: Strong} that secondary research is effective for mapping a
methodological landscape, identifying taxonomies, comparing evaluation measures and
exposing gaps. \textbf{Confidence: Moderate} for any conclusion a review draws about
comparative effectiveness, because a review inherits the dataset limitations and reporting
inconsistencies of the studies it summarises. Reviews of network intrusion detection
repeatedly identify exactly this: variation in datasets, metrics and reproducibility that
prevents the underlying results from being
compared~\cite{ahmad2020network,kocher2021machine,chinnasamy2025deep}.

\subsection{Failure modes}

\begin{itemize}
  \item \textbf{The flow diagram that does not reconcile.} If the numbers in the PRISMA
        flow do not add up, or if no record is excluded at full-text stage, the screening
        was not performed as described. This is checkable by any reader in under a minute,
        and we found an instance in our own seed material (\Cref{sec:seed-audit}).
  \item \textbf{Vote counting.} Reporting that ``seven of nine studies found a positive
        effect'' treats studies of wildly different quality and size as equal votes.
  \item \textbf{Single-database retrieval}, which silently inherits that database's
        publisher coverage.
  \item \textbf{Synthesising incommensurable outcomes} into an apparent consensus.
\end{itemize}

\subsection{Checklist}

\begin{itemize}
  \item Protocol written before search.
  \item Queries and dates recorded verbatim.
  \item Two or more databases plus preprints.
  \item Snowballing performed and reported.
  \item Dual screening with an agreement statistic.
  \item Full-text exclusion reasons recorded.
  \item PRISMA flow that reconciles.
  \item Extraction form piloted.
  \item Limitations of the review distinguished from limitations of the included studies.
\end{itemize}

\section{Family B: Design Science Research}
\label{sec:dsr}

\subsection{Purpose and when to use it}

Design science produces and evaluates an artifact, a model, method, framework,
instantiation, that addresses a class of problem. Use it when the contribution is
something that did not previously exist and the question is whether it works. Much of the
enterprise security corpus is design science whether or not it uses the label: maturity
models, threat-modeling languages, risk frameworks, testbeds and training platforms are
all artifacts~\cite{schlette2021ctisoc2m2,johnson2018meta,rosado2022managing,katsantonis2023cyber}.

\subsection{Protocol}

Peffers et al.'s design-science research methodology gives the six activities that have
become standard~\cite{peffers2007design}, and Hevner et al.'s guidelines give the
evaluation discipline~\cite{hevner2004design}:

\begin{enumerate}
  \item \textbf{Problem identification and motivation.} State the problem class, not the
        instance, and justify why an artifact is the appropriate response.
  \item \textbf{Objectives of a solution.} Define what the artifact must achieve, in terms
        that can be measured. This is where most security design science weakens: the
        objectives are stated qualitatively and the evaluation then cannot fail.
  \item \textbf{Design and development.} Build the artifact and describe its structure at
        a level that permits reimplementation.
  \item \textbf{Demonstration.} Show it working on at least one instance of the problem.
  \item \textbf{Evaluation.} Compare observed behaviour against the objectives from step
        2. Distinguish \emph{formative} evaluation, which improves the artifact, from
        \emph{summative} evaluation, which tests it.
  \item \textbf{Communication.} Report the artifact, the design rationale and the
        evaluation.
\end{enumerate}

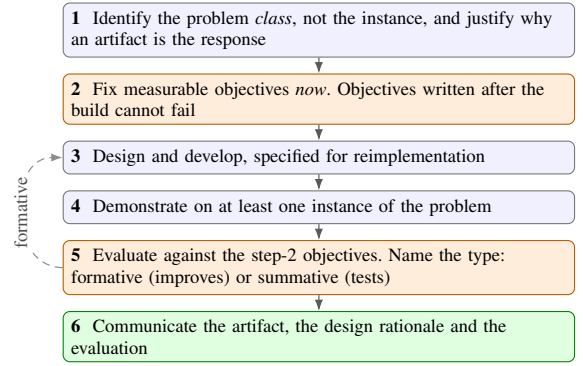
\begin{figure}[t]
\centering
\begin{tikzpicture}[
  node distance=2.1mm,
  mstep/.append style={text width=0.74\columnwidth},
  mgate/.append style={text width=0.74\columnwidth},
  mout/.append style={text width=0.74\columnwidth}]
\node[mstep] (b1) {\sn{1} Identify the problem \emph{class}, not the instance, and justify why an artifact is the response};
\node[mgate, below=of b1] (b2) {\sn{2} Fix measurable objectives \emph{now}. Objectives written after the build cannot fail};
\node[mstep, below=of b2] (b3) {\sn{3} Design and develop, specified for reimplementation};
\node[mstep, below=of b3] (b4) {\sn{4} Demonstrate on at least one instance of the problem};
\node[mgate, below=of b4] (b5) {\sn{5} Evaluate against the step-2 objectives. Name the type: formative (improves) or summative (tests)};
\node[mout, below=of b5] (b6) {\sn{6} Communicate the artifact, the design rationale and the evaluation};
\foreach \a/\b in {b1/b2,b2/b3,b3/b4,b4/b5,b5/b6} \draw[mflow] (\a) -- (\b);
\draw[mback] (b5.west) to[out=180,in=180,looseness=1.3]
  node[mlbl,pos=0.5,rotate=90]{formative} (b3.west);
\end{tikzpicture}
\caption{Family B, after Peffers et al.~\cite{peffers2007design}. The dashed edge is the
formative loop, and drawing it makes the distinction that this family most often loses: an
iteration from evaluation back into design improves the artifact and cannot also test it.
A paper that runs only this loop has demonstrated feasibility, not effectiveness.}
\label{fig:dsr}
\end{figure}

The protocol is illustrated in \Cref{fig:dsr}.

\subsection{Evaluation designs, in increasing strength}

\begin{enumerate}
  \item \emph{Illustrative scenario} --- the artifact is applied to a constructed example.
        Weakest; demonstrates feasibility only.
  \item \emph{Expert evaluation} --- practitioners assess the artifact. Used for the
        CTI-SOC maturity model, evaluated by expert interviews together with a
        prototype~\cite{schlette2021ctisoc2m2}, and for model-based dependency analysis,
        validated through two case studies and expert
        interviews~\cite{jiang2023modelbased}.
  \item \emph{Case application in a real organisation} --- as in the MARISMA-CPS risk
        pattern applied in a smart-hospital case~\cite{rosado2022managing} and the
        model-based security design method assessed against industrial
        cases~\cite{shaked2023modelbased}.
  \item \emph{Comparative evaluation against an existing artifact} on a shared task.
  \item \emph{Controlled experiment} with participants or systems randomised to artifact
        versus baseline. Rare in this corpus, and the strongest available.
\end{enumerate}

\begin{figure}[t]
\centering
\begin{tikzpicture}[
  node distance=1.9mm,
  mstep/.append style={text width=0.68\columnwidth}]
\node[mstep, fill=green!16, draw=green!50!black] (e5)
  {\sn{5} Controlled experiment --- artifact versus baseline, randomised};
\node[mstep, fill=green!8, draw=green!45!black, below=of e5] (e4)
  {\sn{4} Comparative evaluation against an existing artifact on a shared task};
\node[mstep, fill=yellow!14, draw=olive, below=of e4] (e3)
  {\sn{3} Case application in a real organisation};
\node[mstep, fill=orange!14, draw=orange!70!black, below=of e3] (e2)
  {\sn{2} Expert evaluation --- practitioners assess the artifact};
\node[mstep, fill=red!9, draw=red!55!black, below=of e2] (e1)
  {\sn{1} Illustrative scenario --- applied to a constructed example};
\draw[mflow] ($(e1.west)+(-2.4mm,0)$) -- ($(e5.west)+(-2.4mm,0)$)
  node[mlbl, midway, rotate=90]{strength of claim};
\draw[draw=black!45, line width=0.4pt]
  ($(e3.east)+(1.4mm,2.8mm)$) -- ($(e3.east)+(2.6mm,2.8mm)$)
  -- ($(e2.east)+(2.6mm,-2.8mm)$) -- ($(e2.east)+(1.4mm,-2.8mm)$);
\coordinate (brk) at ($(e2.east)!0.5!(e3.east)$);
\node[mnote, right=3.0mm of brk, text width=0.17\columnwidth]
  {where this corpus evaluates};
\end{tikzpicture}
\caption{Family B evaluation designs, weakest at the bottom. A claim of operational
effectiveness needs rung~4 or~5; in our coding, the enterprise security corpus reviewed
here evaluates mostly at rungs~2 and~3, which is why the confidence label for that claim is
\emph{Limited} rather than the label the individual papers' care would otherwise support.}
\label{fig:dsrladder}
\end{figure}
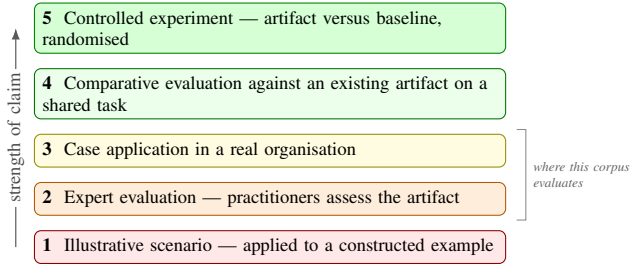

The evaluation ladder is illustrated in \Cref{fig:dsrladder}.

\subsection{What the evidence supports}

\textbf{Confidence: Moderate} that design science is the appropriate vehicle for
framework, model and platform contributions in enterprise security.
\textbf{Confidence: Limited} for claims of operational effectiveness derived from design
science alone, because, in our coding of the corpus (\Cref{sec:review-limits}),
evaluation is predominantly formative, expert-based
or single-case. The decision-support artifact for threat and incident managers is candid
about this: iterative formative evaluation with ten professionals across three financial
institutions indicated potential usefulness, and no summative evaluation was
conducted~\cite{vanderkleij2022developing}.

\subsection{Failure modes}

Objectives written after the artifact; ``evaluation'' that only demonstrates; the artifact
described at a level that prevents reimplementation; expert panels drawn from the authors'
own collaborators; and the absence of any comparison to what practitioners already do,
which is the true baseline.

\subsection{Checklist}

\begin{itemize}
  \item Problem class stated.
  \item Measurable objectives fixed before build.
  \item Artifact specified for reimplementation.
  \item Evaluation type named as formative or summative.
  \item Baseline identified.
  \item Threats to the evaluation's validity stated.
  \item Artifact released (\Cref{sec:reproducibility}).
\end{itemize}

\section{Family C: Qualitative and Socio-Technical Inquiry}
\label{sec:qualitative}

\subsection{Purpose and when to use it}

Qualitative inquiry answers why practitioners act as they do, how work is actually
organised, and what a technical measure is missing. Use it when the phenomenon is
organisational, when the variables are not yet known, or when a quantitative result is
inexplicable. The corpus shows this repeatedly: topic modeling of information-security
journals combined with a Delphi study of corporate chief information security officers
exposed a gap between academic emphasis and practitioner
priority~\cite{dhillon2021information}; interviews with incident-response teams surfaced
communication and collaboration failures between response and security management and
yielded a security-learning model~\cite{ahmad2015case}; a needs assessment across five
Dutch computer security incident response teams organised performance problems into
organisational, team, individual and instrumental
categories~\cite{vanderkleij2017computer}.

Security operations centres are the setting where this family has produced findings that no
technical measurement reaches. A qualitative study of matched and mismatched
centres, interviewing analysts and managers within the same organisations, located the
problems in the relationship between the two groups rather than in either
alone~\cite{kokulu2019matched}, which is a result only obtainable by sampling both sides of
a boundary. A complementary study of analyst challenges and performance metrics documents
what the centre is actually measured on and what that measurement omits~\cite{agyepong2019challenges}.
The methodological point generalises: where a technical metric and a practitioner account
disagree, the disagreement is usually the finding.

\subsection{Reflexive thematic analysis: the protocol}

Braun and Clarke's six phases are the most-used framework and the most-often
misused~\cite{braun2006using}. The authors' own later critical review found that
published applications frequently violate the framework's assumptions, most commonly by
treating themes as things that ``emerge'' and by seeking inter-coder agreement in a design
whose premises do not support it~\cite{braun2023thematic}. The phases:

\begin{enumerate}
  \item \textbf{Familiarisation.} Read the whole corpus actively; write analytic notes.
        Transcribe your own interviews if at all possible.
  \item \textbf{Systematic coding.} Generate codes across the entire dataset, not just the
        parts that interest you. Codes are short, specific and grounded in the data
        (\texttt{escalation-blocked-by-ownership}, not \texttt{communication}).
  \item \textbf{Generating initial themes.} Collate codes into candidate themes. A theme is
        a pattern of shared meaning organised around a central concept, not a topic
        summary.
  \item \textbf{Developing and reviewing themes.} Check candidates against the coded
        extracts and against the whole dataset. Collapse, split or discard.
  \item \textbf{Refining, defining and naming themes.} Write a short definition for each,
        stating its boundary and what it excludes.
  \item \textbf{Writing up.} The analysis is written, not reported; extracts are evidence
        within an argument, not illustration after it.
\end{enumerate}

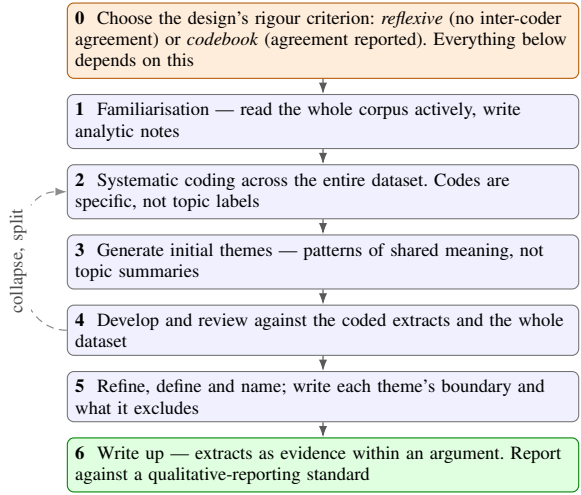
\begin{figure}[t]
\centering
\begin{tikzpicture}[
  node distance=2.0mm,
  mstep/.append style={text width=0.74\columnwidth},
  mgate/.append style={text width=0.74\columnwidth},
  mout/.append style={text width=0.74\columnwidth}]
\node[mgate] (c0) {\sn{0} Choose the design's rigour criterion: \emph{reflexive} (no inter-coder agreement) or \emph{codebook} (agreement reported). Everything below depends on this};
\node[mstep, below=of c0] (c1) {\sn{1} Familiarisation --- read the whole corpus actively, write analytic notes};
\node[mstep, below=of c1] (c2) {\sn{2} Systematic coding across the entire dataset. Codes are specific, not topic labels};
\node[mstep, below=of c2] (c3) {\sn{3} Generate initial themes --- patterns of shared meaning, not topic summaries};
\node[mstep, below=of c3] (c4) {\sn{4} Develop and review against the coded extracts and the whole dataset};
\node[mstep, below=of c4] (c5) {\sn{5} Refine, define and name; write each theme's boundary and what it excludes};
\node[mout, below=of c5] (c6) {\sn{6} Write up --- extracts as evidence within an argument. Report against a qualitative-reporting standard};
\foreach \a/\b in {c0/c1,c1/c2,c2/c3,c3/c4,c4/c5,c5/c6} \draw[mflow] (\a) -- (\b);
\draw[mback] (c4.west) to[out=180,in=180,looseness=1.2]
  node[mlbl,pos=0.5,rotate=90]{collapse, split} (c2.west);
\end{tikzpicture}
\caption{Family C, following Braun and Clarke's six phases~\cite{braun2006using}. Step~0 is
not in the original and is the one this corpus most often gets wrong: the authors' own
critical review found published applications reporting an agreement statistic inside a
reflexive design, whose premises do not support one~\cite{braun2023thematic}. The choice
has to be made before coding, because it determines what counts as rigour afterwards.}
\label{fig:qual}
\end{figure}

The protocol is illustrated in \Cref{fig:qual}.

\subsection{Choosing among the qualitative designs}

\begin{itemize}
  \item \textbf{Semi-structured interviews} when you need practitioner accounts.
        Purposive sampling; an interview guide with open questions and planned probes;
        recruitment continued until additional interviews stop changing the coding.
        ``Saturation'' is the usual word for that stopping rule and it is used loosely:
        Saunders et al. distinguish several incompatible conceptualisations and require a
        study to say which one it operationalised and how it
        judged it reached~\cite{saunders2017saturation}. A sample size defended only by
        the word itself has defended nothing.
        Knowles et al.'s study of the penetration-testing ecosystem interviewed 54
        stakeholders and used the result to argue for standardisation of scoping, delivery
        and reporting~\cite{knowles2016simulated}. Where the practitioner is a developer
        rather than a defender, the same design applies: Naiakshina et al. combined a
        programming task with qualitative interviewing to establish \emph{why} developers
        store passwords insecurely, and found that the omission was usually a matter of
        what the task appeared to ask for rather than of
        knowledge~\cite{naiakshina2017developers}, a mechanism no survey of developer
        knowledge would have surfaced.
  \item \textbf{Case study} when the phenomenon cannot be separated from its context.
        Runeson and H\"ost's guidelines specify the case and unit of analysis, the
        multiple data sources, the chain of evidence and the reporting
        structure~\cite{runeson2008guidelines}; the incident-response case analysis in an
        Australian financial organisation is a worked example~\cite{ahmad2015case}.
  \item \textbf{Cognitive task and cognitive work analysis} when the object of study is
        the decision itself, what the analyst attends to, what they must hold in mind,
        where the load falls. This produced a critical-thinking memory aid for threat and
        incident managers balancing operational context against
        investigation~\cite{vanderkleij2022developing}.
  \item \textbf{Delphi} when you need structured expert consensus on something no dataset
        contains. Okoli and Pawlowski's design guidance covers panel selection, the
        anonymous iterative rounds and the stopping rule~\cite{okoli2004delphi}; a zero-trust
        maturity assessment framework derived its critical success factors this
        way~\cite{yeoh2023zero}.
  \item \textbf{Human-factors mixed methods} when management and employee perspectives
        diverge. In pilot healthcare organisations, combining surveys, interviews and
        telemetry showed that a stronger security culture did not necessarily produce more
        rule-compliant behaviour, and that conflicts among rules and procedures could
        themselves create vulnerabilities~\cite{pollini2021leveraging}.
\end{itemize}

\subsection{Rigour and reliability}
\label{sec:kappa}

Two positions exist and a study must choose one and be consistent. In \emph{reflexive}
thematic analysis, the researcher's interpretation is the instrument, and inter-coder
agreement is not an appropriate criterion; rigour is demonstrated through
reflexivity, an audit trail and analytic depth~\cite{braun2023thematic}. In
\emph{codebook} or content-analytic designs, agreement is appropriate and should be
reported.

Where agreement is reported, Cohen's $\kappa$ corrects observed agreement for chance:
\begin{equation}
\kappa = \frac{p_o - p_e}{1 - p_e}
\label{eq:kappa}
\end{equation}
where $p_o$ is observed agreement and $p_e$ is the agreement expected by chance from the
marginal distributions. Landis and Koch's interpretive bands, slight, fair, moderate, substantial, almost perfect, remain the common
reference~\cite{landis1977measurement}. However, McHugh's account of the statistic's behaviour
demonstrates that these bands are unreliable in skewed
distributions~\cite{mchugh2012interrater}. Two properties matter in practice. First,
$\kappa$ is depressed when one category dominates, which is the normal situation in
security coding; a high $p_o$ with a low $\kappa$ is a prevalence artifact, not
necessarily poor coding. Second, $\kappa$ handles exactly two coders and nominal
categories. For more coders, ordinal categories or missing data, Krippendorff's $\alpha$
is the appropriate measure~\cite{hayes2007answering}:
\begin{equation}
\alpha = 1 - \frac{D_o}{D_e}
\label{eq:alpha}
\end{equation}
with $D_o$ the observed disagreement and $D_e$ the disagreement expected by chance. Report
the statistic, the number of coders, the proportion of the corpus double-coded, and how
disagreements were resolved.

Two reporting standards exist for this family and neither is used in security research as
often as it should be. The Consolidated Criteria for Reporting Qualitative Research (COREQ) is a 32-item checklist for interview and focus-group studies,
organised around the research team, the study design and the analysis, and it exists
because qualitative papers routinely omit the details that let a reader judge the
account, who conducted the interviews, what relationship they had with participants, how
many refused~\cite{tong2007consolidated}. The Standards for Reporting Qualitative Research (SRQR) is the broader instrument, covering
qualitative designs generally rather than interview studies specifically, and is derived
from a synthesis of earlier standards~\cite{obrien2014standards}. Write against one of them
from the start; both are far easier to satisfy prospectively than to retrofit, because
several items concern decisions made before data collection.

\subsection{What the evidence supports}

\textbf{Confidence: Moderate} that organisational coordination, cognitive workload, rule
conflict and temporal variation in behaviour materially affect security outcomes and are
not visible to technical measurement. Self-report bias, small purposive samples,
sector-specific sites and limited transferability remain the standing limitations.

\subsection{Checklist}

\begin{itemize}
  \item Design named and its rigour criterion consistent with it.
  \item Sampling strategy and its rationale.
  \item Recruitment and refusal reported.
  \item Interview guide included.
  \item Coding procedure described.
  \item If agreement is claimed, statistic, coder count and double-coded proportion reported.
  \item Reflexivity statement.
  \item Ethics approval.
  \item Extracts attributed to distinct participants rather than repeatedly to the most articulate one.
\end{itemize}

\section{Family D: Survey, Behavioural and Longitudinal Designs}
\label{sec:survey}

\subsection{Purpose and when to use it}

Surveys estimate how a construct is distributed in a population and what covaries with it.
Use them when the construct is latent, intention to comply, perceived threat severity,
security culture, and when you can define and reach a population. Protection-motivation
and institutional-governance accounts of information-security policy compliance are the
recurring theoretical frames~\cite{hina2019institutional}.

\subsection{Protocol}

\begin{enumerate}
  \item \textbf{Define the population and the sampling frame separately.} The population
        is who the claim is about; the frame is who could actually be sampled. The
        difference between them is the coverage error, and it must be stated.
  \item \textbf{Operationalise constructs} using previously validated scales wherever they
        exist, citing the source. New items require a pilot and a validation step.
  \item \textbf{Determine sample size from the intended analysis}, not from convenience.
        Cohen's power primer gives the conventional effect-size benchmarks and the
        relationship between $\alpha$, power, effect size and $n$~\cite{cohen1992power}.
  \item \textbf{Pilot} with 10--30 respondents from the target population; check item
        comprehension and completion time.
  \item \textbf{Administer}, recording the invitation count, the response rate and the
        completion rate. Test for non-response bias by comparing early and late
        respondents.
  \item \textbf{Assess the measurement model before the structural model.} For PLS-SEM,
        Hair et al. specify the sequence: indicator reliability, internal consistency,
        convergent validity (average variance extracted), discriminant validity
        (heterotrait--monotrait ratio), and only then path coefficients, $R^2$, $f^2$ and
        bootstrapped significance~\cite{hair2019use}.
  \item \textbf{Report} the full measurement model, not only the path diagram. Where the
        survey was administered over the web, which in enterprise security it almost
        always is, report it against CHERRIES, which asks for the things a web survey can
        silently omit: whether the sample was open or closed, how the invitation was
        delivered, whether completeness checks or adaptive questioning were used, and how
        duplicate submissions were prevented~\cite{eysenbach2004improving}. It also fixes
        the ambiguity that makes web-survey response rates incomparable, by distinguishing
        the view rate, the participation rate and the completion rate rather than reporting
        one number called ``the response rate''.
\end{enumerate}

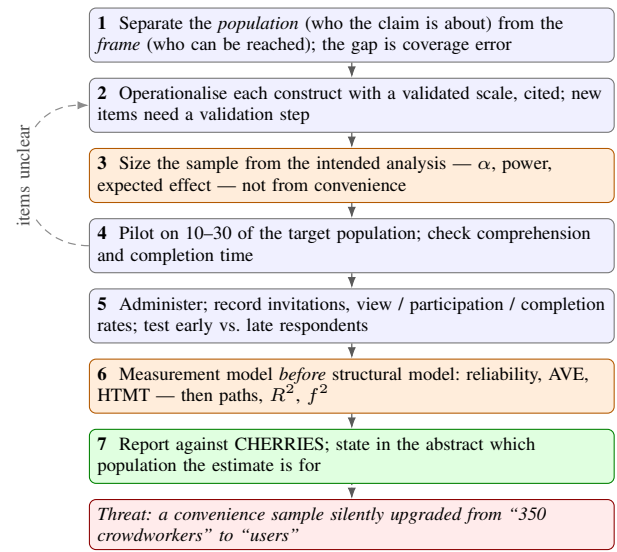
\begin{figure}[t]
\centering
\begin{tikzpicture}[
  node distance=2.0mm,
  mstep/.append style={text width=0.76\columnwidth},
  mgate/.append style={text width=0.76\columnwidth},
  mout/.append style={text width=0.76\columnwidth}]
\node[mstep] (d1) {\sn{1} Separate the \emph{population} (who the claim is about) from the \emph{frame} (who can be reached); the gap is coverage error};
\node[mstep, below=of d1] (d2) {\sn{2} Operationalise each construct with a validated scale, cited; new items need a validation step};
\node[mgate, below=of d2] (d3) {\sn{3} Size the sample from the intended analysis --- $\alpha$, power, expected effect --- not from convenience};
\node[mstep, below=of d3] (d4) {\sn{4} Pilot on 10--30 of the target population; check comprehension and completion time};
\node[mstep, below=of d4] (d5) {\sn{5} Administer; record invitations, view / participation / completion rates; test early vs.\ late respondents};
\node[mgate, below=of d5] (d6) {\sn{6} Measurement model \emph{before} structural model: reliability, AVE, HTMT --- then paths, $R^2$, $f^2$};
\node[mout, below=of d6] (d7) {\sn{7} Report against CHERRIES; state in the abstract which population the estimate is for};
\node[mrisk, below=of d7, text width=0.76\columnwidth] (d8) {Threat: a convenience sample silently upgraded from ``350 crowdworkers'' to ``users''};
\foreach \a/\b in {d1/d2,d2/d3,d3/d4,d4/d5,d5/d6,d6/d7} \draw[mflow] (\a) -- (\b);
\draw[mback] (d4.west) to[out=180,in=180,looseness=1.6]
  node[mlbl,pos=0.5,rotate=90]{items unclear} (d2.west);
\draw[mflow, densely dotted] (d7) -- (d8);
\end{tikzpicture}
\caption{Family D. Step~6 is ordered, not a list: a path coefficient computed on a
measurement model that has not been shown to be reliable and discriminant is a number
about nothing. Step~3 is a gate because a survey underpowered for its intended test cannot
be repaired after collection. The pilot feeds back into operationalisation, which is the
only cheap place to fix a misread item.}
\label{fig:survey}
\end{figure}

The protocol is illustrated in \Cref{fig:survey}.

\subsection{The generalisation problem, and what to do about it}

Security and privacy survey research relies heavily on convenience samples, and the
question of whether those results transfer has been tested directly. Redmiles et al.
compared results obtained from a census-representative panel, from a crowdworker platform
and from a web panel, and found that the answer depends on the measure: some estimates
transfer acceptably and others do not, in ways that are predictable from the
demographics involved~\cite{redmiles2019well}. The practical consequence is not that
convenience samples are unusable. It is that a paper using one must state which population
its estimate is for, and must not silently upgrade ``350 crowdworkers'' into ``users''.

\subsection{Longitudinal and within-person designs}

Cross-sectional surveys identify between-person patterns at one point in time. They cannot
identify within-person patterns, even though most behavioural theory in security is stated
in within-person terms, that a \emph{given} employee complies less when they
neutralise. Cram et al. make this argument and test it, applying an idiographic approach
to neutralisation theory through a four-week experience-sampling study, and finding that
cybersecurity behaviour varies within the same person over
time~\cite{cram2024time}. Where the theory is within-person, a one-time survey is the
wrong instrument regardless of its sample size.

Experience sampling in practice: recruit a smaller panel; prompt several times per week at
semi-random times; keep each prompt under two minutes; expect and report attrition; analyse
with multilevel models that separate within-person from between-person variance.

\subsection{Structured expert elicitation: the Delphi design}

Some questions have no population to sample. ``What are the most effective defences
against social engineering in this sector?'' cannot be answered by asking a representative
sample of employees, because most of them have no basis for an answer; it can only be put
to people who have worked the problem. A one-shot survey of experts is the weak version of
this, because it records opinions without letting them meet each other. Delphi is the
structured version: successive anonymous rounds, each fed by the aggregated responses of
the last, continued until the panel converges or stably fails to.

Campbell's study of social-engineering countermeasures is a usable model of the design in
this setting: 20 information-security practitioners, three rounds, with each round
condensing the previous round's items and returning them for re-rating, producing three
prioritised problem areas and a set of countermeasures the panel
converged on~\cite{campbell2019solutions}. The elements that make such a study reportable
are the ones to copy: the recruitment criterion that defines what ``expert'' meant here,
the panel size and its attrition between rounds, the stopping rule fixed in advance, and
the aggregation statistic used to decide that consensus had been reached. Anonymity between
panellists is the point of the design, not an administrative detail, it is what stops the
most senior participant from setting the answer in round one.

Delphi licenses a claim about \emph{expert consensus} and nothing stronger. It does not
establish that the countermeasures work; it establishes that practitioners with stated
qualifications agreed that they do. Papers in this family fail most often by reporting the
consensus as though it were an effectiveness result.

\subsection{What the evidence supports}

\textbf{Confidence: Moderate} for survey designs as instruments for latent organisational
constructs, conditional on measurement-model reporting.
\textbf{Confidence: Limited} for the within-person finding, which rests on a single
four-week study, though the methodological argument for the design is independent of that
study's results.

\subsection{Checklist}

\begin{itemize}
  \item Population and frame distinguished.
  \item Scales cited or validated.
  \item Power analysis reported.
  \item Pilot conducted.
  \item View, participation and completion rates reported separately.
  \item Non-response bias tested.
  \item Measurement model reported in full before structural results.
  \item The population the estimate generalises to stated in the abstract.
  \item CHERRIES items addressed for a web-administered instrument. For a Delphi study: the expertise criterion, the panel size and per-round attrition, the pre-specified stopping rule, and the consensus statistic.
\end{itemize}

\section{Family E: Threat Modeling, Attack Graphs and Quantitative Risk}
\label{sec:threat}

\subsection{Purpose and when to use it}

Threat modeling connects technical weaknesses to enterprise risk decisions. Use it when
the question concerns which attack paths exist in a system, what they would cost an
attacker, and where a control would be most effective. We begin with Spoofing, Tampering, Repudiation, Information disclosure, Denial of service and Elevation of privilege (STRIDE), the most widely used enumeration.

\subsection{STRIDE: the protocol}

\begin{enumerate}
  \item \textbf{Decompose} the system into processes, data stores, data flows,
        external entities and trust boundaries. A data-flow diagram is the working
        artifact.
  \item \textbf{Enumerate} threats per element against the six categories: spoofing,
        tampering, repudiation, information disclosure, denial of service, elevation of
        privilege. Not every category applies to every element type, and the
        element-to-category mapping is what makes the method systematic rather than
        free-form.
  \item \textbf{Assess} each threat for impact and likelihood.
  \item \textbf{Plan countermeasures} and record the residual risk of each accepted threat.
\end{enumerate}

\begin{figure}[t]
\centering
\begin{tikzpicture}[
  node distance=2.0mm,
  mstep/.append style={text width=0.76\columnwidth},
  mgate/.append style={text width=0.76\columnwidth},
  mout/.append style={text width=0.76\columnwidth},
  mrisk/.append style={text width=0.76\columnwidth}]
\node[mstep] (e1) {\sn{1} Decompose: processes, data stores, data flows, external entities --- and draw the trust boundaries. The data-flow diagram is the working artifact};
\node[mgate, below=of e1] (e2) {\sn{2} Per element, enumerate only the \emph{applicable} STRIDE categories. The element-to-category mapping is what makes this systematic rather than free-form};
\node[mrisk, below=of e2] (e2r) {Blind spot: threats that exist only in the \emph{interaction} between components};
\node[mgate, below=of e2r] (e3) {\sn{3} Do the threats compose into multi-step paths?};
\node[mstep, below=of e3] (e4) {\sn{4a} No: assess impact $\times$ likelihood per threat};
\node[mstep, below=of e4] (e5) {\sn{4b} Yes: lift to a path representation --- attack tree / graph, staged attack-simulation processes, or a probabilistic architecture model};
\node[mstep, below=of e5] (e6) {\sn{5} Plan countermeasures against the ranked threats};
\node[mout, below=of e6] (e7) {\sn{6} Residual-risk register: every accepted threat, with who accepted it};
\foreach \a/\b in {e1/e2,e2/e2r,e2r/e3,e3/e4,e4/e5,e5/e6,e6/e7} \draw[mflow] (\a) -- (\b);
% Routed rectangularly rather than as a wide bezier: over a span of five nodes the
% curve's control points push the picture past \columnwidth.
\coordinate (er) at ($(e1.east)+(2.6mm,0)$);
\draw[mback] (e5.east) -- (e5.east -| er)
  -- node[mlbl,rotate=270]{new elements} (er) -- (e1.east);
\end{tikzpicture}
\caption{Family E. Step~2 is a gate because free-form threat brainstorming is not STRIDE;
the systematicity lives entirely in the element-to-category mapping. Step~3 is the decision
this family most often skips: enumeration answers ``what can go wrong here'', and only a
path representation answers ``what can an adversary chain''. A model refined at 4b usually
exposes elements missing from the data-flow diagram, so the loop back to step~1 is expected, not a sign
of a bad first pass.}
\label{fig:stride}
\end{figure}
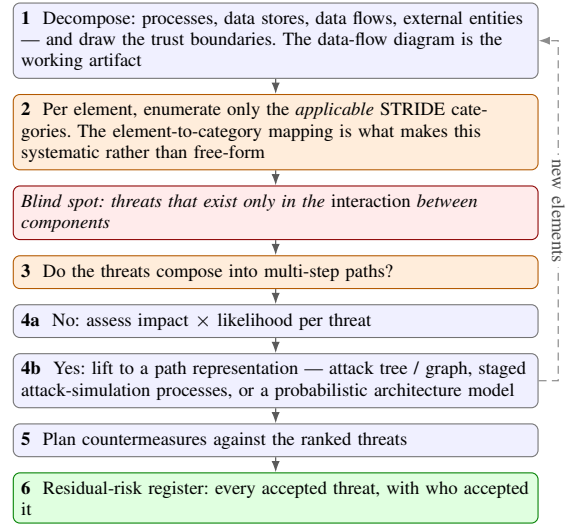

The protocol is illustrated in \Cref{fig:stride}.

Applied to a synchrophasor cyber-physical testbed with data-flow diagrams, STRIDE was
characterised as lightweight and effective for component-level analysis, with the explicit
caveat that threats arising only from interactions \emph{between} components can be
missed~\cite{khan2017stridebased}. A container-ecosystem study used the same method across
repositories, image registries, containers and hosts, and then surveyed countermeasure
strengths and weaknesses, which is the correct pairing, since threat enumeration without
mitigation analysis produces a list nobody can act
on~\cite{wong2023security}.

\subsection{Risk-centric and graph-based extensions}

STRIDE enumerates; it does not compose. Multi-step attacks require a representation with
paths.

\begin{itemize}
  \item \textbf{Process for Attack Simulation and Threat Analysis} integrates threat modeling with the development lifecycle across
        seven stages and aligns countermeasures with business
        impact~\cite{ucedavelez2015risk}.
  \item \textbf{Integrated software- and attack-centric modeling} combines system models,
        component models, attack trees, attack graphs and Common Vulnerability Scoring System (CVSS) scoring; a railway
        communication network is the worked case~\cite{potteiger2016software}.
  \item \textbf{Attack-graph meta-languages} formalise attack-step dependencies so that
        attack sequences and time-to-compromise can be simulated rather than
        enumerated by hand~\cite{johnson2018meta}.
  \item \textbf{Probabilistic architecture models.} CySeMoL estimates the probability of
        successful attack in an enterprise architecture by probabilistic inference, and
        was validated by comparison against security professionals' judgement and case
        studies~\cite{sommestad2013cyber}.
\end{itemize}

\subsection{Reducing dependence on expert construction}

The standing weakness of this family is that the model is built by hand by an expert, so
it scales poorly and inherits that expert's assumptions. Two directions address it.
Malcoda gathers data automatically from deployed security products, constructs a Bayesian
directed acyclic graph and computes risk probabilities per asset and vulnerability; in a
virtual enterprise network it completed assessments quickly, produced probabilities that
tracked intrusion-detection alerts, and had computational complexity no greater than
comparable models~\cite{sato2025malcoda}. Dynamic multilevel risk modeling pursues the
same goal for large distributed systems using security metrics, though the specific
datasets and validation environment were not reported, which is itself the reason we
label it lower~\cite{palko2023cyber}. Security assurance metrics work supplies the
measurement layer these models need~\cite{wen2022developing}.

\subsection{Decision-theoretic and network-analytic extensions}

Two further quantitative treatments answer questions that enumeration and probability
alone do not.

\textbf{Decision-theoretic investment modeling} asks not which paths exist but how much to
spend and where. Formulating enterprise resilience through expected utility, with $x$ the
defensive investment and $y$ the recovery investment, gives an objective of the form
\begin{equation}
\begin{split}
EU(x,y) = {}& \bigl(1-p(x)\bigr)\,U(W-x-y) \\
            &+ p(x)\,U\bigl(W-x-y-L(y)\bigr)
\end{split}
\label{eq:eu}
\end{equation}
where $p(x)$ is the breach probability under defensive spend $x$, $L(y)$ the residual loss
after recovery spend $y$, $W$ the initial endowment and $U$ a utility function encoding the
organisation's risk attitude. An expected-resilience framework built on this basis models
defence and recovery investment jointly and derives allocation
strategies~\cite{dong2024building}. The methodological requirement is the same as for any
model in this family: state where $p(\cdot)$, $L(\cdot)$ and the risk-attitude parameter
come from, and run sensitivity analysis over them, because the optimum is a function of
assumptions rather than of measurements.

\textbf{Network analysis of technique taxonomies} treats a framework such as MITRE ATT\&CK
as a graph and asks which techniques are structurally central to attack paths. Applying
semantic network analysis to enterprise ATT\&CK techniques identifies the ones that
recur across pathways, credential access and valid-account abuse being the usual
result, and thereby offers a defensible ordering for control
prioritisation~\cite{graham2025enhancing}. Betweenness centrality is the standard
instrument,
\begin{equation}
C_B(v) = \sum_{s \neq v \neq t \in V} \frac{\sigma_{st}(v)}{\sigma_{st}}
\label{eq:betweenness}
\end{equation}
with $\sigma_{st}$ the number of shortest paths from $s$ to $t$ and $\sigma_{st}(v)$ the
number of those passing through $v$. Two cautions apply. The centrality of a technique in
the taxonomy reflects how the taxonomy was written, not how often adversaries use the
technique; and a graph built from documented procedures inherits the reporting bias of the
threat intelligence that produced them. Weighting edges by observed incident frequency,
where such data exist, converts a structural claim into an empirical one.

A historical benchmark is worth keeping in view. Ortalo et al. monitored a large
operational system for nearly two years, modelling privilege graphs and estimating the
effort required of an attacker~\cite{ortalo1999experimenting}. Longitudinal operational
measurement of this kind remains uncommon, and it is what most modern risk models are
implicitly claiming to approximate.

\subsection{What the evidence supports}

\textbf{Confidence: Moderate} that threat modeling and probabilistic or graph-based
analysis usefully structure enterprise risk. Confidence in generalisation is reduced by
case-specific assumptions, dependence on expert input, limited validation against real
incidents, and threat conditions that change faster than models are rebuilt.

\subsection{Failure modes}

Treating a CVSS base score as a risk estimate; models whose parameters are elicited from
the same experts who then validate them; attack graphs whose state space is pruned in
undocumented ways; and probability outputs reported without sensitivity analysis over the
inputs that are least well known.

\subsection{Checklist}

\begin{itemize}
  \item System decomposition and trust boundaries shown.
  \item Method named and its scope limit stated.
  \item Parameter sources identified and separated from validators.
  \item Sensitivity analysis over uncertain inputs.
  \item Validation against something other than the authors' own judgement.
  \item Residual risk recorded for accepted threats.
\end{itemize}

\section{Family F: Simulation and Model-Based Analysis}
\label{sec:simulation}

\subsection{Purpose and when to use it}

Simulation is the method of choice where direct experimentation would disrupt a critical
process or require exercising a dangerous capability. Use it to explore cascading
consequences, to run sensitivity analyses that no live system would tolerate, and to
compare defensive configurations at a scale that field measurement cannot reach.

\subsection{The main formalisms}

\begin{itemize}
  \item \textbf{Stochastic Petri nets} model the interaction between adversary and defence
        as concurrent stochastic processes, and support optimisation of design parameters.
        Applied to a modernised electrical grid, this identified intrusion-detection
        intervals and redundancy levels that maximise mean time to
        failure~\cite{mitchell2016modeling}.
  \item \textbf{System dynamics with sensitivity analysis} measures how covariances among
        observed variables change before and after an intervention. Experiments using IEEE
        14-bus and 300-bus grid models together with the Tennessee Eastman chemical process
        demonstrated efficiency, scalability and cross-sector applicability within the
        tested scenarios~\cite{genge2015system}.
  \item \textbf{Integrated cyber-physical simulation with attacker--defender games.} The
        SURE platform combines cyber and physical models, attack models, operational
        scenarios and game-theoretic reasoning, applied to smart transportation including
        traffic-signal tampering, resilient sensor selection and denial of
        service~\cite{koutsoukos2018sure}.
  \item \textbf{Dependency modeling for cascade analysis.} Model-based cybersecurity
        analysis partitions cyber and cyber-physical functional dependencies to model
        cascading effects, instantiated on power-grid models and a municipal grid and
        checked against expert interviews~\cite{jiang2023modelbased}.
\end{itemize}

\subsection{Protocol}

\begin{enumerate}
  \item State the question as a comparison between configurations or a sensitivity over
        parameters. Simulation answers ``what changes if'' well and ``what is true'' badly.
  \item Specify the model: state variables, transitions, stochastic assumptions, time
        semantics.
  \item Identify every parameter and its source. Separate measured parameters, estimated
        parameters and assumed parameters, and label them as such in the paper.
  \item Verify the implementation (does it compute the model) before validating the model
        (does the model correspond to reality). These are different activities and both are
        needed.
  \item Establish run length and replication count from the variance of the output, and
        report confidence intervals over replications rather than single runs.
  \item Run sensitivity analysis over the assumed parameters. If a conclusion inverts
        within the plausible range of an assumed parameter, that is the finding.
  \item Validate against something external: an operational incident record, expert
        judgement, or a physical testbed.
\end{enumerate}

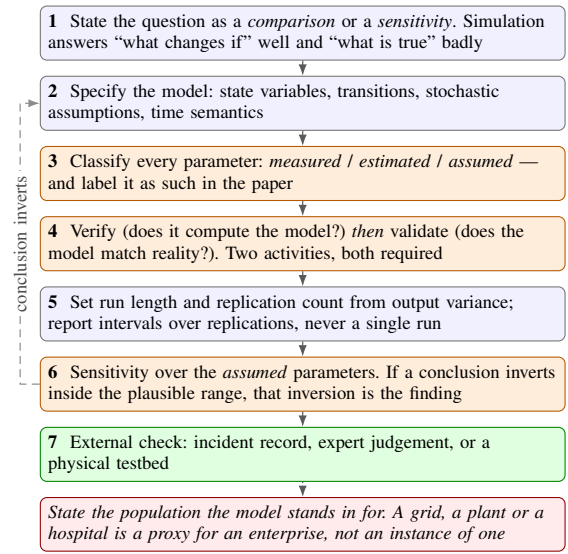
\begin{figure}[t]
\centering
\begin{tikzpicture}[
  node distance=2.0mm,
  mstep/.append style={text width=0.76\columnwidth},
  mgate/.append style={text width=0.76\columnwidth},
  mout/.append style={text width=0.76\columnwidth},
  mrisk/.append style={text width=0.76\columnwidth}]
\node[mstep] (f1) {\sn{1} State the question as a \emph{comparison} or a \emph{sensitivity}. Simulation answers ``what changes if'' well and ``what is true'' badly};
\node[mstep, below=of f1] (f2) {\sn{2} Specify the model: state variables, transitions, stochastic assumptions, time semantics};
\node[mgate, below=of f2] (f3) {\sn{3} Classify every parameter: \emph{measured} / \emph{estimated} / \emph{assumed} --- and label it as such in the paper};
\node[mgate, below=of f3] (f4) {\sn{4} Verify (does it compute the model?) \emph{then} validate (does the model match reality?). Two activities, both required};
\node[mstep, below=of f4] (f5) {\sn{5} Set run length and replication count from output variance; report intervals over replications, never a single run};
\node[mgate, below=of f5] (f6) {\sn{6} Sensitivity over the \emph{assumed} parameters. If a conclusion inverts inside the plausible range, that inversion is the finding};
\node[mout, below=of f6] (f7) {\sn{7} External check: incident record, expert judgement, or a physical testbed};
\node[mrisk, below=of f7] (f8) {State the population the model stands in for. A grid, a plant or a hospital is a proxy for an enterprise, not an instance of one};
\foreach \a/\b in {f1/f2,f2/f3,f3/f4,f4/f5,f5/f6,f6/f7} \draw[mflow] (\a) -- (\b);
\draw[mflow, densely dotted] (f7) -- (f8);
\coordinate (fl) at ($(f2.west)-(2.6mm,0)$);
\draw[mback] (f6.west) -- (f6.west -| fl)
  -- node[mlbl,rotate=90]{conclusion inverts} (fl) -- (f2.west);
\end{tikzpicture}
\caption{Family F. Three of the seven steps are gates, which is the distinguishing feature
of this family: a simulation always produces numbers, so the discipline has to come from
the protocol rather than from the possibility of failure. Step~3 is the one most often
omitted, an unlabelled parameter table lets a reader assume every value was measured, and
it is what makes step~6 meaningful, since sensitivity analysis is only informative over the
parameters that were assumed.}
\label{fig:sim}
\end{figure}

The protocol is illustrated in \Cref{fig:sim}.

\subsection{What the evidence supports}

\textbf{Confidence: Moderate} for the methodological usefulness of simulation in this
setting. \textbf{Confidence: Limited} for any prediction outside the modelled
infrastructure or the modelled scenario set. Model fidelity is the binding constraint:
simulated attacks, dependencies and operational responses need not reproduce the behaviour
of a real organisation, whose responses are made by people under time pressure. The studies
here examined grids, transport, chemical process and a hospital, which are proxies for, not
instances of, a general enterprise population.

\subsection{Checklist}

\begin{itemize}
  \item Model specification complete enough to reimplement.
  \item Parameters classified by source.
  \item Verification distinct from validation.
  \item Replication count justified.
  \item Confidence intervals over replications.
  \item Sensitivity analysis on assumed parameters.
  \item External validation attempt.
  \item Explicit statement of the population the model stands in for.
\end{itemize}

\section{Family G: Machine-Learning Experimentation}
\label{sec:ml}

\subsection{Purpose and when to use it}

This is the dominant methodology for detection research and the one with the most
thoroughly documented pathologies. Use it when the question is whether a model can
discriminate a class of events, under stated conditions, at an operating point that is
usable.

\subsection{The baseline protocol}

\begin{enumerate}
  \item \textbf{Define the operational task} before choosing data: what is detected, at
        what latency, with what consequence for a false positive and for a false negative.
        The consequence asymmetry determines the metric and the operating point, so it
        cannot be decided afterwards.
  \item \textbf{Select data} and state its provenance, collection period, labelling
        procedure and label quality. Benchmarks in wide use here include KDD~99, NSL-KDD,
        CIC-IDS2017, CIC-IDS2018 and UNSW-NB15; each has documented artifacts, and using
        one is a decision to be justified rather than a default. CIC-IDS2017 is the
        instructive case, because two independent re-examinations found defects in it and
        both found that the defects change results. Engelen et al. traced errors to the
        traffic generation, the attack execution and the labelling, and reported that
        correcting them alters the measured detection performance~\cite{engelen2021troubleshooting};
        Lanvin et al. quantified the difference the corrections make and found it
        substantial enough to affect comparisons between
        detectors~\cite{lanvin2023errors}. Before adopting any benchmark, search for work
        that has audited it, and state which version or corrected release you used.
  \item \textbf{Split before you preprocess.} Fit every transformation, normalisation,
        feature selection, encoding, dimensionality reduction, on the training partition
        only, then apply to validation and test. Fitting on the full dataset leaks test
        information into training and inflates every downstream number.
  \item \textbf{Handle class imbalance explicitly, and only inside training folds.}
        Oversampling applied before the split fabricates near-duplicates of test points
        into the training set. This is the most common single defect in the detection
        literature.
  \item \textbf{Choose the split to match the deployment question.} Random $k$-fold
        answers ``can the model separate these classes''. It does not answer ``will the
        model work next month'', because random splitting places future samples in the
        training set. For that question the split must be temporal
        (\Cref{sec:temporal}).
  \item \textbf{Tune hyperparameters on a validation partition}, never on test, and report
        the search space and budget.
  \item \textbf{Report the full confusion matrix} plus precision, recall, $F_1$, and a
        threshold-free curve. Report training and inference cost, which determine
        deployability.
  \item \textbf{Compare against a meaningful baseline}: a simple model, and where possible
        the rule or heuristic actually in production.
  \item \textbf{Repeat with different seeds} and report variance. A single-run difference
        of half a point is not a result.
\end{enumerate}

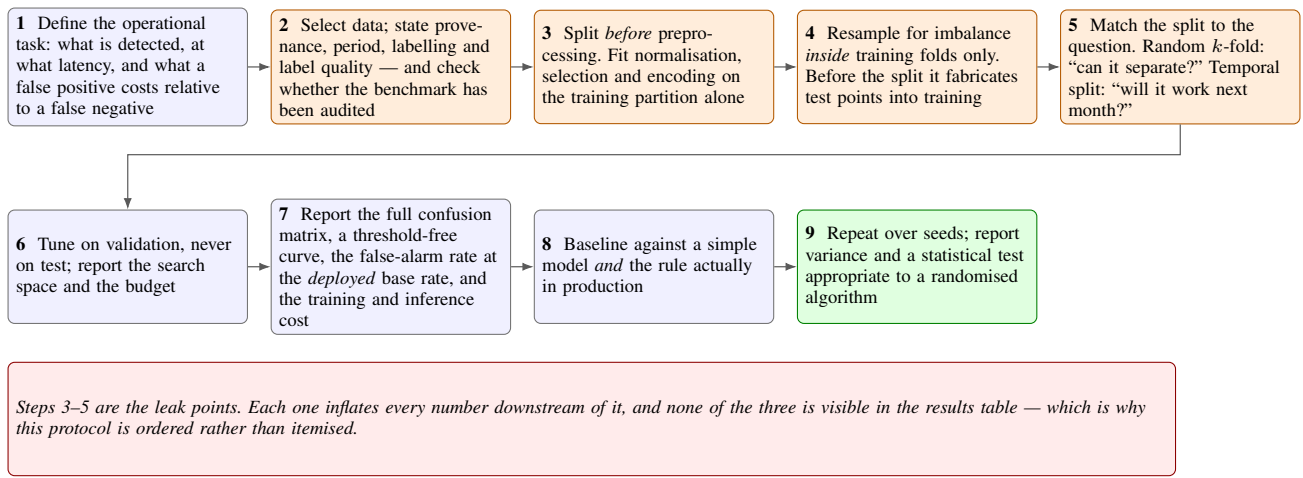
\begin{figure*}[t]
\centering
\begin{tikzpicture}[
  node distance=3mm,
  mstep/.append style={text width=0.163\textwidth, minimum height=15mm},
  mgate/.append style={text width=0.163\textwidth, minimum height=15mm},
  mout/.append style={text width=0.163\textwidth, minimum height=15mm}]
\node[mstep] (g1) {\sn{1} Define the operational task: what is detected, at what latency, and what a false positive costs relative to a false negative};
\node[mgate, right=of g1] (g2) {\sn{2} Select data; state provenance, period, labelling and label quality --- and check whether the benchmark has been audited};
\node[mgate, right=of g2] (g3) {\sn{3} Split \emph{before} preprocessing. Fit normalisation, selection and encoding on the training partition alone};
\node[mgate, right=of g3] (g4) {\sn{4} Resample for imbalance \emph{inside} training folds only. Before the split it fabricates test points into training};
\node[mgate, right=of g4] (g5) {\sn{5} Match the split to the question. Random $k$-fold: ``can it separate?'' Temporal split: ``will it work next month?''};
\node[mstep, below=11mm of g1.south west, anchor=north west] (g6) {\sn{6} Tune on validation, never on test; report the search space and the budget};
\node[mstep, right=of g6] (g7) {\sn{7} Report the full confusion matrix, a threshold-free curve, the false-alarm rate at the \emph{deployed} base rate, and the training and inference cost};
\node[mstep, right=of g7] (g8) {\sn{8} Baseline against a simple model \emph{and} the rule actually in production};
\node[mout, right=of g8] (g9) {\sn{9} Repeat over seeds; report variance and a statistical test appropriate to a randomised algorithm};
\node[mrisk, below=5mm of g6.south west, anchor=north west, text width=0.84\textwidth]
  (grisk) {Steps 3--5 are the leak points. Each one inflates every number downstream of it, and none of the three is visible in the results table --- which is why this protocol is ordered rather than itemised.};
\foreach \a/\b in {g1/g2,g2/g3,g3/g4,g4/g5,g6/g7,g7/g8,g8/g9}
  \draw[mflow] (\a) -- (\b);
\draw[mflow] (g5.south) -- ++(0,-4mm) -| (g6.north);
\end{tikzpicture}
\caption{Family G. The order matters more than the content. Steps~3--5 concern only where
a boundary is drawn in the data, cost nothing to get right, and are the defects most often
found in published detection work; a paper that fits a scaler on the whole dataset before
splitting reports numbers that no amount of later rigour repairs. Step~5 is a gate rather
than a step because it is not a methodological preference: random and temporal splits
answer different questions, and only one of them is the question a deployment asks.}
\label{fig:mlproto}
\end{figure*}

The protocol is illustrated in \Cref{fig:mlproto}.

\subsection{Metrics, and why accuracy is the wrong one}

With $TP$, $FP$, $TN$, $FN$ from the confusion matrix:
\begin{equation}
\mathrm{Precision} = \frac{TP}{TP+FP}, \qquad
\mathrm{Recall} = \frac{TP}{TP+FN}
\label{eq:pr}
\end{equation}
\begin{equation}
F_1 = 2\cdot\frac{\mathrm{Precision}\cdot \mathrm{Recall}}{\mathrm{Precision}+\mathrm{Recall}}
\label{eq:f1}
\end{equation}

Benign traffic vastly outnumbers attack traffic, so accuracy is dominated by the majority
class and a trivial always-benign classifier scores well. The deeper problem is the
base-rate fallacy, stated for intrusion detection by Axelsson: because the prior
probability of an intrusion in any given event is extremely small, the quantity an analyst
actually experiences is the Bayesian detection rate
\begin{equation}
P(I \mid A) = \frac{P(A \mid I)\,P(I)}{P(A \mid I)\,P(I) + P(A \mid \neg I)\,P(\neg I)}
\label{eq:bayes}
\end{equation}
which remains low even for a detector with an excellent true-positive rate, unless the
false-alarm rate $P(A \mid \neg I)$ is driven extraordinarily
low~\cite{axelsson2000baserate}. A paper reporting 99\% recall without reporting the
false-alarm rate at the deployed base rate has not addressed the question an operator
has. Sommer and Paxson's analysis of why machine learning is harder for intrusion detection
than for other domains, the cost asymmetry, the semantic gap between an anomaly and an
attack, the difficulty of obtaining ground truth, remains the reference for framing
this~\cite{sommer2010outside}.

Where the deployment cares about ranking under heavy imbalance, precision--recall curves
are more informative than receiver-operating-characteristic curves, because the latter's
false-positive-rate axis is dominated by the large negative class and compresses exactly
the region of interest.

\subsection{Temporal bias, spatial bias and drift}
\label{sec:temporal}

Two biases invalidate a large fraction of published detection results. \emph{Temporal
bias} occurs when training data contains samples that post-date test samples.
\emph{Spatial bias} occurs when the ratio of malicious to benign samples in the test set
does not correspond to any realistic deployment. TESSERACT formalised both, gave
constraints that a sound evaluation must satisfy, and introduced a
time-aware metric so that classifiers can be compared over a deployment
period rather than at a point~\cite{pendlebury2018tesseract}; the extended treatment
develops the framework further~\cite{kan2024tesseract}. Any study whose claim concerns
future performance owes the reader a temporally consistent split.

Concept drift is the underlying phenomenon: the joint distribution of features and labels
changes over time, so a model degrades after deployment. The general adaptation literature
classifies drift and the responses to it~\cite{gama2014survey}. In security specifically,
conformal-evaluation approaches identify samples that a deployed classifier should reject
rather than classify; the Transcend line of work revisited these methods, analysed their
statistical basis and reported the conditions under which they
hold~\cite{barbero2022transcending}.

Studying drift requires a dataset with a usable time axis, which most security benchmarks
do not have, a single snapshot cannot exhibit drift no matter how it is split. Purpose-built
longitudinal benchmarks are the enabling instrument here: LAMDA is an Android malware
benchmark assembled specifically so that concept drift can be measured, with samples
spanning multiple years and timestamps that support temporally consistent
splitting~\cite{haque2025lamda}. When designing a drift study, choose the dataset for its
time coverage first; no evaluation protocol can recover a time axis the data does not
carry.

\subsection{Pitfalls with names}

Arp et al. catalogue ten recurring pitfalls in machine learning for computer security,
demonstrate their prevalence in published work, and quantify how much reported performance
they account for~\cite{arp2020dos}. They include sampling bias, label inaccuracy, data
snooping, spurious correlations, biased parameter selection, inappropriate baselines,
inappropriate performance measures, base-rate neglect, lab-only evaluation and
inappropriate threat models. The corresponding earlier statement for malware experiments
specifically, correct dataset construction, correct handling of goodware, transparency
about what was executed, is Rossow et al.'s prudent
practices~\cite{rossow2012prudent}. Reading both before designing an experiment is a
better use of a week than any amount of additional tuning.

Two further concerns apply when the model is itself the object of attack. Adversarial
machine learning has a decade of accumulated methodology on how evaluations against an
adaptive adversary must be constructed to be meaningful~\cite{biggio2018wild}. And where a
paper claims explainability, the explanation method itself needs evaluating: Warnecke et
al. define criteria, descriptive accuracy, sparsity, stability, efficiency,
robustness, and apply them to explanation methods in security
settings~\cite{warnecke2020evaluating}. Explainable-artificial-intelligence
intrusion-detection work is evaluated against those criteria rather than against the
plausibility of the produced explanation~\cite{mohale2025evaluating}.

\subsection{Comparing randomised systems: the statistics step~9 requires}

Step~9 says ``report variance''. What it means in practice is that a comparison between
two randomised procedures is a statistical question and has to be answered with a
statistical test, not by comparing two means. Arcuri and Briand set out the practical
version of this for software engineering: run enough repetitions, use a non-parametric
test, the Mann--Whitney U test, because the outcome distributions are typically neither
normal nor equally dispersed, report a standardised effect size such as the Vargha--Delaney
$\hat{A}_{12}$ alongside the $p$-value, and correct for multiple comparisons when several
configurations are compared at once~\cite{arcuri2011practical}. All four apply unchanged
to a detector compared across seeds, and the effect size is the one most often missing:
with enough repetitions any difference becomes significant, and only the effect size says
whether it matters.

Klees et al. make the same argument for fuzzing and make it concretely, by re-running
published evaluations under a corrected protocol. Their findings, that results depend
heavily on the seed corpus, that the timeout chosen changes which tool wins, that
crash de-duplication by stack hash inflates bug counts, and that most published
comparisons used too few trials to support their conclusion, are specific to fuzzing, but
the underlying failure is not~\cite{klees2018evaluating}. It is the practice of reporting
one run of a stochastic process as though it were a measurement. Where a security
experiment involves any randomised component, seed, initialisation, sampling, scheduling,
or an attack generator, the same three requirements hold: multiple trials, a test that
matches the distribution, and an effect size.

\subsection{What the evidence supports}

\textbf{Confidence: Strong} that comparative experimentation with transparent
preprocessing, splitting and metric reporting is the appropriate design for detection
questions. \textbf{Confidence: Conflicting} for any claim that one algorithm family
dominates in enterprise deployment; \Cref{sec:contradiction} analyses this directly.
\textbf{Confidence: Limited} for transferring any benchmark result to a live enterprise
network, because, in our coding, the corpus evaluates almost entirely on benchmark,
simulated or laboratory
data rather than on continuously collected enterprise
populations~\cite{gamage2020deep,zhang2022comparative,figueiredo2023deep,shone2018deep,javaid2016deep,yin2017deep}.

\subsection{Checklist}

\begin{itemize}
  \item Operational task and cost asymmetry stated.
  \item Data provenance and labelling described.
  \item Preprocessing fitted inside the training partition only.
  \item Resampling inside folds only.
  \item Split type justified against the deployment question.
  \item Temporal consistency where the claim is about the future.
  \item Hyperparameter search space and budget reported.
  \item Full confusion matrix.
  \item False-alarm rate at a realistic base rate.
  \item Training and inference cost.
  \item Baseline including the incumbent.
  \item Multiple seeds with a non-parametric test and an effect size.
  \item Code and configuration released.
\end{itemize}

This family has no reporting standard of its own, which is part of why the same defects
recur. The nearest usable instrument comes from clinical prediction modelling: the Transparent Reporting of a multivariable prediction model for Individual Prognosis Or Diagnosis (TRIPOD) is a
22-item checklist for reporting a multivariable prediction model, and its items map onto
detection research almost directly, how participants (here, samples) were selected, how
predictors were defined and measured, how missing data were handled, how the model was
specified, how performance was assessed, and what happened at
validation~\cite{collins2015transparent}. It is not a security instrument and its
vocabulary has to be translated, but a detection paper that can answer all 22 items has
answered the questions that make a model assessable.

\section{Family H: Evaluating Language-Model and Agentic Systems}
\label{sec:llm}

\subsection{Why this needs its own methodology}

Language-model-based systems break several assumptions the previous section relies on.
The training corpus is unknown, so a benchmark may already be inside it. The output is
free text, so scoring requires a judge. The system is often agentic, so behaviour depends
on the environment and on the number of attempts allowed. Each of these creates a distinct
validity threat, and a synthesis of pitfalls in language-model security research documents
how frequently they go unaddressed~\cite{evertz2025chasing}.

\subsection{Contamination and construct validity}

Benchmark contamination is the check to run before any other. If a public benchmark predates the
model, a strong score may measure recall rather than capability. Mitigations: construct or
hold out tasks after the model's cutoff; perturb tasks so that memorised solutions do not
transfer; report the model version and date; and treat public-benchmark results as an
upper bound. A related failure appears in log analysis, where passive content in the
context window can steer the analysis, context contamination that a
conventional accuracy measure does not
detect~\cite{karanjai2026context}.

\subsection{Language models as judges}

Using a model to score another model's output is now routine and needs its own validity
argument. A systematisation of security concerns in the judge setting catalogues the
attack surface and the failure
modes~\cite{masoud2026security}. More pointedly, a large-scale evaluation of judge models
found high reliability, the judge agrees with itself, without corresponding
validity, meaning consistent scores that do not track the quality being
claimed~\cite{norman2026reliability}. Reliability is therefore not evidence of validity,
and reporting only agreement between judge runs is not an evaluation. Minimum practice:
calibrate the judge against human labels on a sample, report that agreement with a chance-
corrected statistic (\Cref{eq:kappa} or \Cref{eq:alpha}), randomise presentation order,
and check for position and verbosity effects.

\subsection{Agentic security systems}

Benchmarking agents that act, scanning, exploiting, remediating, is harder again, because
the environment is part of the measurement and success is path-dependent. A direct
treatment of why this is hard identifies the ways an agent benchmark can flatter its
subject~\cite{abdelnabi2026measuring}. Concrete studies show what a careful design looks
like: automated web penetration testing evaluated against a defined capability
ladder~\cite{wu2025autopt}; and a comparison of artificial-intelligence agents against
cybersecurity professionals in real-world penetration testing, which is the design that
answers the question practitioners actually ask~\cite{lin2025comparing}. Where the output
is a detection artifact rather than an action, the artifact can be evaluated on its own
terms, as in the assessment of model-generated detection
rules~\cite{bertiger2025evaluating}. Systematic review of algorithmic red-teaming
methodology organises this emerging space~\cite{srivastava2026systematic}.

\subsection{Protocol additions}

Beyond the machine-learning checklist: state the model, version and access date; state the
sampling temperature and the number of attempts permitted per task, and report
pass-at-$k$ rather than best-of-$n$ presented as a single attempt; hold out a
contamination-controlled task set; if a judge is used, report its calibration against human
labels; describe the environment as precisely as a testbed paper would; and report cost per
task, because an agent that succeeds at arbitrary expense has not solved an operational
problem.

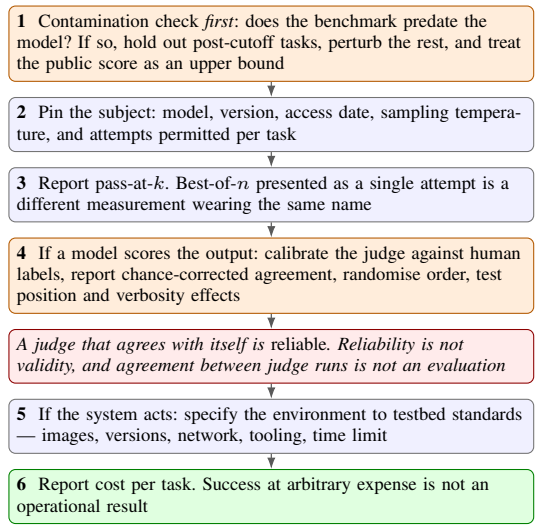
\begin{figure}[t]
\centering
\begin{tikzpicture}[
  node distance=2.0mm,
  mstep/.append style={text width=0.76\columnwidth},
  mgate/.append style={text width=0.76\columnwidth},
  mout/.append style={text width=0.76\columnwidth},
  mrisk/.append style={text width=0.76\columnwidth}]
\node[mgate] (h1) {\sn{1} Contamination check \emph{first}: does the benchmark predate the model? If so, hold out post-cutoff tasks, perturb the rest, and treat the public score as an upper bound};
\node[mstep, below=of h1] (h2) {\sn{2} Pin the subject: model, version, access date, sampling temperature, and attempts permitted per task};
\node[mstep, below=of h2] (h3) {\sn{3} Report pass-at-$k$. Best-of-$n$ presented as a single attempt is a different measurement wearing the same name};
\node[mgate, below=of h3] (h4) {\sn{4} If a model scores the output: calibrate the judge against human labels, report chance-corrected agreement, randomise order, test position and verbosity effects};
\node[mrisk, below=of h4] (h5) {A judge that agrees with itself is \emph{reliable}. Reliability is not validity, and agreement between judge runs is not an evaluation};
\node[mstep, below=of h5] (h6) {\sn{5} If the system acts: specify the environment to testbed standards --- images, versions, network, tooling, time limit};
\node[mout, below=of h6] (h7) {\sn{6} Report cost per task. Success at arbitrary expense is not an operational result};
\foreach \a/\b in {h1/h2,h2/h3,h3/h4,h4/h5,h5/h6,h6/h7} \draw[mflow] (\a) -- (\b);
\end{tikzpicture}
\caption{Family H, as additions to the Family G protocol rather than a replacement for it.
Step~1 is first because it is disqualifying: if the benchmark is in the training corpus,
nothing measured afterwards is about capability. Steps~4 and~6 are the two that most often
go unreported, and they fail in opposite directions, an uncalibrated judge makes a result
look better than it is, while an unreported cost makes an unusable system look deployable.}
\label{fig:llm}
\end{figure}

The protocol is illustrated in \Cref{fig:llm}.

\subsection{What the evidence supports}

\textbf{Confidence: Moderate} that contamination control, judge calibration and
environment specification are necessary for a defensible claim in this family.
\textbf{Confidence: Limited} for any current claim about operational capability of agentic
security systems, because the evaluation methodology is younger than the systems it
measures.

\subsection{Checklist}

\begin{itemize}
  \item Model, version and access date reported.
  \item Sampling temperature and attempts permitted per task stated.
  \item pass-at-$k$ reported, not best-of-$n$ presented as a single attempt.
  \item Contamination check run first; public-benchmark scores treated as upper bounds.
  \item Judge calibrated against human labels, with a chance-corrected agreement statistic.
  \item Presentation order randomised; position and verbosity effects tested.
  \item Environment specified to testbed standards (\Cref{sec:infrastructure}).
  \item Cost per task reported.
\end{itemize}

\section{Family I: Measurement, Telemetry and Detection Efficacy}
\label{sec:measurement}

\subsection{Purpose and when to use it}

This family asks what is observable at all. It precedes detection research logically: a
detector cannot find what the telemetry does not record, and a large part of the apparent
gap between laboratory and production performance lives here rather than in the model.

\subsection{Instruments and findings}

Logging is the substrate. A measurement of the detection efficacy of modern security
logging standards asks what fraction of adversary behaviour the standards' recommended
event sets actually make visible~\cite{holeman2026collection}, which is the right question
and one rarely asked before a detector is proposed.

Endpoint detection and response products are evaluated publicly through adversary-emulation
exercises, and those results can themselves be analysed as data. A study decoding the MITRE
Engenuity ATT\&CK Enterprise evaluation examined endpoint-detection performance across
participants and drew out what the published results do and do not
establish~\cite{shen2024decoding}. Reading a vendor evaluation as a measurement instrument,
with its own construct validity, is a transferable methodological move.

Detector evaluation is also sensitive to how the evaluation set is constructed. For
graph-based lateral-movement detection, an analysis of fair and realistic performance
evaluation shows how the choice of graph, of anomaly injection and of base rate determines
the reported result~\cite{larroche2026fair}. The same lesson appears in industrial
intrusion detection, where a systematisation of evaluation practice found the field's
comparisons hard to interpret for want of shared
protocol~\cite{lamberts2023evaluations}.

Where the behaviour to be measured is the adversary's rather than the defender's, a
honeypot is the instrument, and its configuration is not a deployment detail but the
independent variable. Lupia et al. exposed an industrial-control honeynet emulating
operational-technology devices to the Internet for three months and analysed the resulting
interactions along three axes, level of interaction, origin, and interaction or attack
pattern, reporting how configuration choices affect both the attractiveness of the
honeynet and the behaviour it captures~\cite{lupia2023ics}. That is the methodological
finding worth carrying: a honeypot measurement is conditional on the bait, so a paper must
report the emulated device set, the exposure period and the network placement before any
frequency it reports can be interpreted.

Threat-intelligence quality is the analogous measurement problem on the intelligence side:
what is in a feed, how timely is it, and does it support the decisions it is bought
for~\cite{tounsi2018survey,martins2022generating}. Comparative analysis of six incident-response
intelligence formats identified 18 core concepts and concluded that organisations may
reasonably combine formats by use case~\cite{schlette2021comparative}.

\subsection{Protocol}

\begin{enumerate}
  \item Define the observable of interest as a behaviour, not as a product feature.
  \item Establish ground truth independently of the system being measured, emulation with
        a known plan, an injected campaign, or a labelled incident record.
  \item Measure coverage first (what fraction of the behaviour appears in telemetry at
        all), then efficacy (what fraction of what appears is detected).
  \item Report the configuration exactly: agent versions, rule sets, log sources enabled,
        retention.
  \item Separate detection from alerting from response; a behaviour can be logged,
        detected, and still never reach an analyst.
  \item Report against Strengthening the Reporting of Observational Studies in Epidemiology (STROBE). Everything in this family is an observational study, and
        STROBE is the reporting standard for one: it asks for the design, the setting and
        the dates, the eligibility criteria and how participants, here, hosts, sessions
        or events, were selected, how each variable was measured, how much data were
        missing, and what the study cannot rule out~\cite{vonelm2007strengthening}. Its
        items translate to telemetry work with almost no adaptation, and it is the
        instrument that stops a measurement paper from reporting a number without its
        denominator.
\end{enumerate}

\begin{figure}[t]
\centering
\begin{tikzpicture}[
  node distance=2.0mm,
  mstep/.append style={text width=0.76\columnwidth},
  mgate/.append style={text width=0.76\columnwidth},
  mout/.append style={text width=0.76\columnwidth},
  mrisk/.append style={text width=0.76\columnwidth}]
\node[mstep] (i1) {\sn{1} Define the observable as a \emph{behaviour}, not as a product feature};
\node[mgate, below=of i1] (i2) {\sn{2} Establish ground truth \emph{independently} of the system being measured: an emulation plan, an injected campaign, or a labelled incident record};
\node[mstep, below=of i2] (i3) {\sn{3a} Coverage: what fraction of the behaviour reaches the telemetry at all?};
\node[mstep, below=of i3] (i4) {\sn{3b} Efficacy: of what reaches it, what fraction is detected?};
\node[mstep, below=of i4] (i5) {\sn{4} Report the configuration exactly: agent versions, rule sets, log sources enabled, retention};
\node[mgate, below=of i5] (i6) {\sn{5} Separate logged from detected from alerted from actioned. A behaviour can be all of the first three and still reach no analyst};
\node[mout, below=of i6] (i7) {\sn{6} Report against STROBE, with the denominator for every rate};
\node[mrisk, below=of i7] (i8) {A detector cannot find what the telemetry never recorded. Measuring efficacy without measuring coverage attributes a logging gap to the model};
\foreach \a/\b in {i1/i2,i2/i3,i3/i4,i4/i5,i5/i6,i6/i7} \draw[mflow] (\a) -- (\b);
\draw[mflow, densely dotted] (i7) -- (i8);
\end{tikzpicture}
\caption{Family I. The ordering of 3a before 3b is the whole point of the family: coverage
and efficacy are different quantities with different denominators, and a study that
reports only the second cannot tell a weak detector from an absent log source. Step~2 is a
gate because ground truth derived from the system under measurement makes every subsequent
number circular.}
\label{fig:meas}
\end{figure}
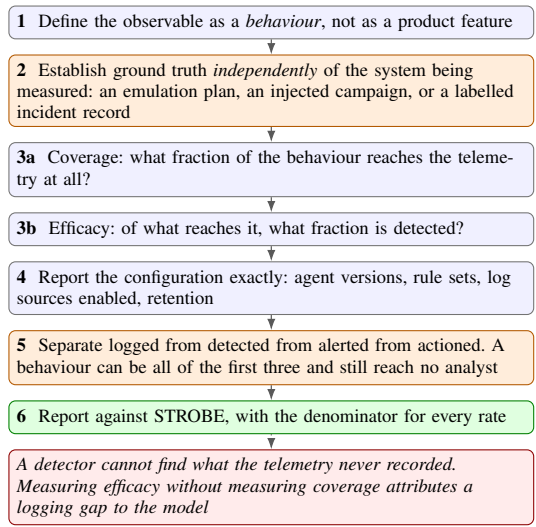

The protocol is illustrated in \Cref{fig:meas}.

\subsection{Measuring consequences: the event-study design}

Telemetry measures what happened inside the system. A second measurement question, what
the incident cost, cannot be answered from telemetry at all, and has its own established
design. The event study, borrowed from financial economics, takes a set of publicly dated
events, estimates each firm's normal return from a pre-event window, and tests whether the
return in a narrow window around the announcement departs from it. Campbell et al.\ applied
this to publicly announced information-security breaches, using stock-market reaction as the
measured consequence~\cite{campbell2003economic}.

The design is worth knowing because it answers a question researchers frequently assert
rather than measure, and because its constraints are instructive. It requires publicly
traded firms, publicly dated announcements, and an event window short enough that no other
material news falls inside it; it measures the market's expectation of cost, not the cost;
and it inherits a selection problem, since breaches that are never announced are never in
the sample. Any modern replication has a further difficulty: mandatory-disclosure regimes
have changed both which breaches are announced and how quickly, so the sampling frame is
not comparable across regulatory periods. State the frame and the window, and claim the
expectation rather than the cost.

\subsection{What the evidence supports}

\textbf{Confidence: Moderate} that telemetry coverage is a first-order determinant of
detection outcomes and is under-measured relative to model performance.

\subsection{Checklist}

\begin{itemize}
  \item Observable defined as a behaviour, not a product feature.
  \item Ground truth established independently of the system measured.
  \item Coverage measured before efficacy, each with its denominator.
  \item Exact configuration reported: agent versions, rule sets, log sources, retention.
  \item Logged, detected, alerted and actioned separated.
  \item Reported against STROBE.
  \item For a honeypot: emulated device set, exposure period and network placement stated.
\end{itemize}

\section{Family J: Testbeds, Cyber Ranges and Adversary Emulation}
\label{sec:infrastructure}

\subsection{Purpose and when to use it}

Experimental infrastructure is what makes a security experiment both safe and repeatable.
Use it whenever the experiment involves executing attacks, when production access is
impossible, or when a result must be reproducible by someone else.

\subsection{The design space}

A systematic treatment of cyber ranges and security testbeds organises the space by
scenario, function, tool and architecture, and remains the reference
map~\cite{yamin2020cyber}. Recent work moves along three axes:

\begin{itemize}
  \item \textbf{Reproducibility.} The Gotham testbed provides a reproducible environment
        for security experiments and dataset generation, so that a dataset carries its
        generating configuration with it~\cite{saezdecamara2024gotham}. Container-native
        testbeds pursue the same goal for general cybersecurity
        experimentation~\cite{bitzki2026netsecbed}.
  \item \textbf{Scale and federation.} Federating testbeds across institutions addresses
        the topology and heterogeneity that a single laboratory cannot
        provide~\cite{dean2026federated}.
  \item \textbf{Scenario generation.} Building exercise scenarios by hand is the dominant
        cost in range operation, and automated scenario generation targets exactly
        that~\cite{skandylas2026automated}; the cyber-range design framework integrates
        architecture and lifecycle with educational theory and addresses preparation cost,
        assessment weakness and reuse~\cite{katsantonis2023cyber}.
\end{itemize}

\subsection{Adversary emulation as a measurement instrument}

An emulation plan, an ordered sequence of adversary behaviours mapped to a common
technique taxonomy, turns red teaming from a demonstration into a repeatable measurement.
The plan is written first, executed identically across configurations, and scored against
what the defensive stack produced. This is what makes the results of endpoint evaluations
comparable at all~\cite{shen2024decoding}, and it is the mechanism by which a range
exercise yields data rather than an anecdote.

Penetration testing as a research method needs the same discipline. A review of penetration
testing research covers tools, vulnerability classes and automation
trends~\cite{alhamed2023systematic}, and a mixed-method audit of a distributed firewall
shows the reporting form: scanning plus testing, severity classification, and a risk report
that a defender can act on~\cite{tudosi2023research}. The stakeholder study
of the assessment ecosystem is the counterweight, documenting ambiguity in scoping and
delivery that undermines comparability across engagements~\cite{knowles2016simulated}.

\subsection{Studying real adversaries in controlled infrastructure}

Emulation substitutes a researcher for the adversary, which bounds what it can establish:
it measures the defensive stack's response to behaviours the researcher chose. A distinct
design keeps the infrastructure under the researcher's control but lets real offenders
supply the behaviour. Ricaldi et al.\ describe the construction of an Access-as-a-Service
platform modelled on illicit marketplaces, instrumented with manipulable market signals and
a honeypot back end, so that offender target selection and subsequent actions can be
observed and the influence of those signals assessed~\cite{ricaldi2025experimental}. The
paper is published as an experimental design put to the community for critique before the
run, which is itself the practice this family should adopt more widely.

Designs of this kind buy ecological validity at a real price, and the price is ethical
rather than technical. The researcher operates infrastructure that facilitates offending,
cannot obtain consent from the population under study, and must decide in advance what to
do about data on third-party victims. Anyone contemplating one should settle the
containment boundary, the data-minimisation rule and the institutional and legal approvals
before building anything (\Cref{sec:ethics}), and should expect those constraints, not the
engineering, to determine what the study can observe.

\subsection{Protocol}

\begin{enumerate}
  \item Specify the environment as code. A testbed described in prose is not reproducible.
  \item State the fidelity claim: which properties of the real environment the testbed
        reproduces, and, more importantly, which it does not.
  \item Isolate. Document the containment boundary and how live-network egress is
        prevented.
  \item Write the emulation plan before running it, mapped to a technique taxonomy.
  \item Instrument the environment independently of the systems under test.
  \item Publish the configuration, the plan and the resulting data together.
\end{enumerate}

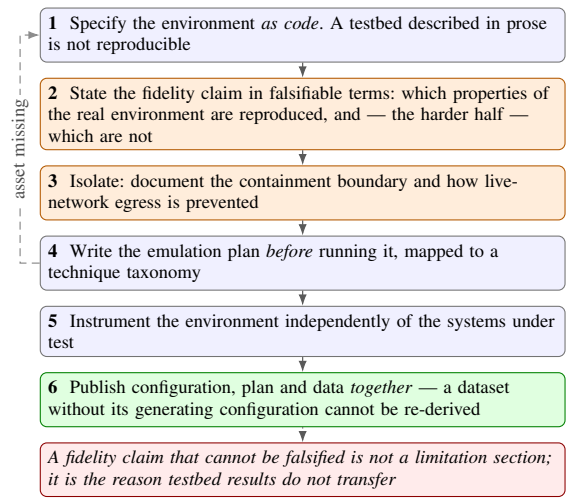
\begin{figure}[t]
\centering
\begin{tikzpicture}[
  node distance=2.0mm,
  mstep/.append style={text width=0.76\columnwidth},
  mgate/.append style={text width=0.76\columnwidth},
  mout/.append style={text width=0.76\columnwidth},
  mrisk/.append style={text width=0.76\columnwidth}]
\node[mstep] (j1) {\sn{1} Specify the environment \emph{as code}. A testbed described in prose is not reproducible};
\node[mgate, below=of j1] (j2) {\sn{2} State the fidelity claim in falsifiable terms: which properties of the real environment are reproduced, and --- the harder half --- which are not};
\node[mgate, below=of j2] (j3) {\sn{3} Isolate: document the containment boundary and how live-network egress is prevented};
\node[mstep, below=of j3] (j4) {\sn{4} Write the emulation plan \emph{before} running it, mapped to a technique taxonomy};
\node[mstep, below=of j4] (j5) {\sn{5} Instrument the environment independently of the systems under test};
\node[mout, below=of j5] (j6) {\sn{6} Publish configuration, plan and data \emph{together} --- a dataset without its generating configuration cannot be re-derived};
\node[mrisk, below=of j6] (j7) {A fidelity claim that cannot be falsified is not a limitation section; it is the reason testbed results do not transfer};
\foreach \a/\b in {j1/j2,j2/j3,j3/j4,j4/j5,j5/j6} \draw[mflow] (\a) -- (\b);
\draw[mflow, densely dotted] (j6) -- (j7);
\coordinate (jl) at ($(j1.west)-(2.6mm,0)$);
\draw[mback] (j4.west) -- (j4.west -| jl)
  -- node[mlbl,rotate=90]{asset missing} (jl) -- (j1.west);
\end{tikzpicture}
\caption{Family J. Steps~2 and~3 are gates for opposite reasons: an unfalsifiable fidelity
claim makes the result untransferable, and an undocumented containment boundary makes the
experiment unsafe to repeat. Writing the emulation plan at step~4 routinely reveals assets
the environment does not have, which is why the specification at step~1 has to be
re-openable rather than fixed once.}
\label{fig:testbed}
\end{figure}

The protocol is illustrated in \Cref{fig:testbed}.

\subsection{What the evidence supports}

\textbf{Confidence: Moderate} that ranges and testbeds are effective research environments
for controlled security experimentation and training. \textbf{Confidence: Limited} for
transferring a testbed result to production, because fidelity claims are seldom stated in
falsifiable terms.

\subsection{Checklist}

\begin{itemize}
  \item Environment specified as code.
  \item Fidelity claim stated falsifiably, including what is not reproduced.
  \item Containment boundary documented; live-network egress prevented.
  \item Emulation plan written before execution and mapped to a technique taxonomy.
  \item Instrumentation independent of the systems under test.
  \item Configuration, plan and data published together.
\end{itemize}

\section{Family K: Field Experiments, Training and Awareness Evaluation}
\label{sec:training}

\subsection{Purpose and when to use it}

This family answers whether an intervention changes what people do. It is the only family
that licenses a causal claim about behaviour, and it requires randomisation to do so.

\subsection{Field experiments}

The canonical enterprise instance is the phishing simulation, and the methodological
question is what the simulation is for. A field experiment on learning through
post-simulation feedback compared embedded and non-embedded feedback and reported effects
on subsequent behaviour~\cite{yin2026learning}. The design elements that make such a study
interpretable: assignment at a unit that prevents contamination between conditions
(commonly the team or department rather than the individual, since colleagues talk); a
pre-registered primary outcome; an intention-to-treat analysis; and an ethics approval that
covers deception of employees.

Report against CONSORT. It is the reporting standard for a parallel-group randomised trial,
and the enterprise field experiment is one: its 25 items ask how the allocation sequence was
generated, how it was concealed until assignment, who was blinded, how many units were
assigned and how many were analysed, and why anyone dropped out~\cite{schulz2010consort}.
The flow diagram it requires, assessed, assigned, received, followed up, analysed, is the
single most useful artifact a phishing-simulation study can publish, because it makes
visible the employees who left the company mid-study, the department that was excluded
after the fact, and the difference between the units randomised and the units analysed.
Wohlin et al.'s treatment of experimentation in software engineering supplies the design
vocabulary underneath it, experiment scoping, planning, operation, analysis and
presentation, with the standard threat classification of conclusion, internal, construct
and external validity, in a form closer to the conditions of a security experiment than the
clinical literature is~\cite{wohlin2012experimentation}.

\subsection{Training and range-based education}

Adaptive training begins from trainee needs, assesses performance continuously and raises
difficulty progressively; implemented on a cyber-range platform for smart-shipping
personnel, it was evaluated through performance outcomes, knowledge acquisition, retention
and incident-response capability~\cite{hatzivasilis2020modern}. Game-based approaches
replace instruction with decision-making; the CyberCIEGE evaluation reported improved
engagement and application of concepts in simulated
environments~\cite{cone2007video}. Both are formative evaluations.

Two recent studies show what the summative version looks like, and both are worth copying
structurally rather than topically. Glas et al.\ re-designed an existing cyber-range
exercise for tier-one analysts in a commercial security-operations centre around
instructional-design principles, structured feedback, scaffolding, competitive
elements, and evaluated the re-design in a randomised controlled trial with 144
participants drawn from cybersecurity courses at two universities, assigned to the original
or the revised learning-management system. They report better learning experience and
shorter training time with equivalent knowledge outcomes~\cite{glas2026ger}. The design
point is the comparator: the control condition was the exercise the organisation was
already running, so the study measures the increment attributable to the re-design rather
than the effect of training versus nothing.

Lee et al.\ address the other standing weakness of this literature, that training is
evaluated away from the job it is meant to change. Their intervention runs gamified phishing
training inside employees' ordinary work context over several weeks rather than in a
dedicated session, and a randomised field experiment found better phishing detection and
avoidance in the treatment group than in the control; they also model a psychological
mechanism, mindfulness of phishing, rather than reporting the effect
alone~\cite{lee2025nothing}. Proposing a mechanism and testing a path model is what
distinguishes a study that explains an effect from one that only records it, and it is the
part most awareness evaluations omit.

\subsection{Awareness programme evaluation}

Measuring whether employees completed training measures compliance with a training policy,
not awareness. A framework derived from a review of 32 papers adapts four indicators from
the European Literacy Policy Network, impact, sustainability, accessibility and
monitoring, to make awareness evaluation systematic and
repeatable~\cite{chaudhary2022developing}. Small and medium-sized enterprises are
identified as an evidence-poor context with respect to awareness needs, resources,
expertise and incentives~\cite{chaudhary2023quest}.

\subsection{Human-subjects methodology in security}

Empirical methods in usable security have themselves been reviewed systematically, with
attention to how risk is represented to participants, a design choice that determines
whether participants behave as they would under real
stakes~\cite{distler2021systematic}. The generalisation caution
of \Cref{sec:survey} applies with full force here~\cite{redmiles2019well}.

\subsection{What the evidence supports}

\textbf{Confidence: Moderate} that ranges, adaptive progression and interactive exercises
improve engagement and short-term performance indicators. \textbf{Confidence: Limited} for
long-term effectiveness: sustained retention, organisational behaviour change and
scalability are not consistently evaluated in this corpus.

\subsection{Checklist}

\begin{itemize}
  \item Randomisation unit stated and contamination addressed.
  \item Allocation concealment described.
  \item Primary outcome pre-specified.
  \item Intention-to-treat analysis.
  \item CONSORT flow reconciling units randomised with units analysed.
  \item Ethics approval covering any deception.
  \item Debriefing procedure.
  \item Outcome measured as behaviour rather than completion.
  \item Comparator stated as the incumbent practice where one exists.
  \item Follow-up at a stated interval.
  \item Effect size with an interval, not only a $p$-value.
\end{itemize}

\begin{figure}[t]
\centering
\begin{tikzpicture}[
  node distance=2.0mm,
  mstep/.append style={text width=0.76\columnwidth},
  mgate/.append style={text width=0.76\columnwidth},
  mout/.append style={text width=0.76\columnwidth},
  mrisk/.append style={text width=0.76\columnwidth}]
\node[mstep] (k1) {\sn{1} State the behaviour the intervention is meant to change --- not the knowledge, and not the completion rate};
\node[mgate, below=of k1] (k2) {\sn{2} Choose the randomisation unit to prevent contamination. Colleagues talk, so the team or department is usually the unit, not the individual};
\node[mgate, below=of k2] (k3) {\sn{3} Name the comparator. ``Versus nothing'' and ``versus the training already running'' are different claims};
\node[mstep, below=of k3] (k4) {\sn{4} Pre-register the primary outcome, the follow-up interval, and the analysis --- before assignment};
\node[mgate, below=of k4] (k5) {\sn{5} Ethics approval covering deception of employees, plus a debriefing procedure};
\node[mstep, below=of k5] (k6) {\sn{6} Assign, deliver, and measure behaviour at the pre-registered interval. Analyse by intention to treat};
\node[mout, below=of k6] (k7) {\sn{7} Report against CONSORT: allocation, concealment, the flow from assigned to analysed, and effect sizes with intervals};
\node[mrisk, below=of k7] (k8) {Measuring who completed the training measures compliance with a training policy. It is not an outcome};
\foreach \a/\b in {k1/k2,k2/k3,k3/k4,k4/k5,k5/k6,k6/k7} \draw[mflow] (\a) -- (\b);
\draw[mflow, densely dotted] (k7) -- (k8);
\end{tikzpicture}
\caption{Family K. This is the only family in the review that licenses a causal claim about
behaviour, and every gate exists to protect that licence. Step~2 protects internal validity,
step~3 decides what the effect is an effect \emph{of}, and step~5 is a gate in the literal
sense: a phishing simulation deceives employees, and without approval and debriefing the
study should not run. Step~4 must precede assignment, which is why it cannot be added
during analysis.}
\label{fig:field}
\end{figure}
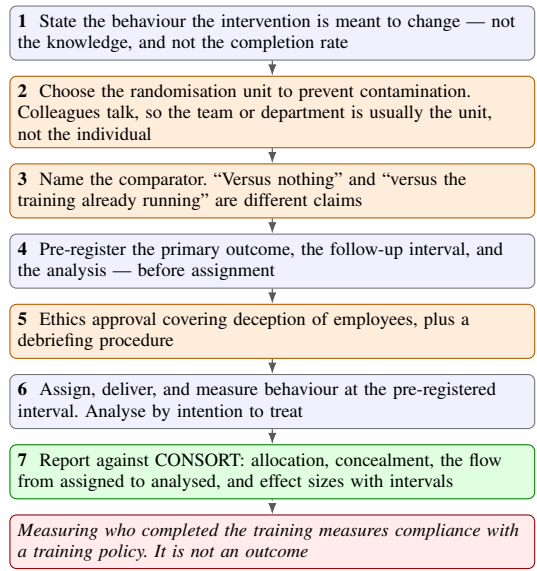

The protocol is illustrated in \Cref{fig:field}.

\section{Operations, Incident Response and Maturity}
\label{sec:operations}

Operational research cuts across the families above and deserves separate treatment
because its evaluation problem is distinctive: the object of study is a capability, not a
component.

Threat-intelligence research identifies a persistently technology-centred discourse and
argues that operationalisation requires attention to processes, practitioners, artifacts,
organisational strategy and sharing practice~\cite{ainslie2023cyberthreat}. The CTI-SOC2M2
maturity model links intelligence integration to security-operations service maturity and
was evaluated through expert interviews and a
prototype~\cite{schlette2021ctisoc2m2}. Practice-oriented incident-response guidance
emphasises policy, roles, documented procedures, threat modeling, risk assessment, regular
exercises, self-assessment, third-party assessment and independent
audit~\cite{nelson2025incident}, aligned to the current cybersecurity
framework~\cite{standards2024nist}. Incident-response team research stresses repeatable
process, response speed, learning from incidents and information
sharing~\cite{ruefle2014computer}.

The methodological lesson is that maturity must be assessed as a capability system rather
than reduced to a single score. \textbf{Confidence: Moderate} for the value of structured,
repeatable, intelligence-informed process. \textbf{Confidence: Limited} for
cross-organisational comparison, because the literature has not converged on validated
measures of incident-response effectiveness and organisational resources and commitment
vary widely.

For researchers, this implies a specific design: pair a maturity instrument with an
independent outcome measure. A maturity level that correlates with nothing observable is a
description of documentation, not of capability.

\subsection{Process models as research objects: the forensic case}

Incident response and digital forensics share a methodological problem the rest of this
review does not have: the process itself is the artifact under scrutiny, and it may have to
survive examination in court. That raises the standard of what counts as a validated
process, and the digital-forensics literature is where the standard has been worked out.

The instructive line of work is the harmonisation effort. Valjarevic and Venter observed
that the field had accumulated many partially overlapping investigation process models, no
two of which decomposed the work the same way, and proposed a harmonised model built by
reconciling the existing ones, an exercise in method consolidation rather than method
invention~\cite{valjarevic2012harmonised}. The extended treatment develops the model
comprehensively and, more importantly for a methodologist, states the criteria by which
such a model should be judged: completeness against the phases the field recognises, and a
correspondence between the model's steps and the evidential requirements they
serve~\cite{valjarevic2015comprehensive}. A researcher proposing any process model, for
response, for triage, for threat hunting, owes the same two things, and most proposals
supply neither.

Hargreaves et al.\ take the layer below. Rather than modelling the investigation, they model
the \emph{tool}: an abstract model of what a digital-forensic analysis tool does, given as a
foundation for systematic error-mitigation analysis, so that the places a tool can be wrong
can be enumerated rather than discovered by
accident~\cite{hargreaves2024abstract}. This is a generalisable move and an under-used one.
Where a field's results depend on tooling, as detection research depends on parsers,
feature extractors and labelling scripts, an explicit model of the tool is what turns
``the tool has bugs'' from a caveat into an analysable error budget.

\subsection{Prioritising organisational change under disagreement}

Capability programmes fail on ordering as often as on content: an organisation cannot do
33 things at once, and which to do first is a judgement no single stakeholder holds. Pigola
and Meirelles treat that ordering as the research question. Working with a panel of 29
professionals, they elicit the criteria that govern zero-trust adoption using a fuzzy Delphi
procedure, which admits that expert judgements are imprecise rather than forcing them to a
point, and then weight the criteria with CRITIC, which derives importance from the
correlation structure among the criteria instead of from further expert opinion. They
report four dimensions, culture, operations and processes, compliance, and
investments, across 33 criteria, and they note that the resulting rank can reverse under
changing conditions~\cite{pigola2025zero}.

That last observation is the methodological content, and it generalises past zero trust.
A multi-criteria weighting is a snapshot of a preference structure; reporting the ranking
without testing its stability presents a contingent ordering as a finding. Any study using
this family of methods should report the sensitivity of its ranking to the weights, in the
same spirit as the sensitivity analysis \Cref{fig:sim} requires of a simulation. The
pairing of an opinion-based elicitation step with a correlation-based weighting step is
also worth noting on its own: it is a deliberate attempt to keep the panel from setting
both the criteria and their importance.

\begin{figure}[t]
\centering
\begin{tikzpicture}[
  node distance=2.0mm,
  mstep/.append style={text width=0.76\columnwidth},
  mgate/.append style={text width=0.76\columnwidth},
  mout/.append style={text width=0.76\columnwidth},
  mrisk/.append style={text width=0.76\columnwidth}]
\node[mstep] (p1) {\sn{1} Enumerate the models the proposal would replace, and say where each one decomposes the work differently};
\node[mgate, below=of p1] (p2) {\sn{2} Reconcile before inventing. Is every phase the existing models recognise accounted for --- kept, merged, or explicitly rejected?};
\node[mstep, below=of p2] (p3) {\sn{3} Map each step to the requirement it serves: evidential, operational, or regulatory. A step serving none is decoration};
\node[mgate, below=of p3] (p4) {\sn{4} Model the tools the process depends on, so that where a tool can be wrong is enumerated rather than discovered};
\node[mstep, below=of p4] (p5) {\sn{5} Instantiate on a real case, not on an illustrative one};
\node[mout, below=of p5] (p6) {\sn{6} Pair the model with an \emph{independent} outcome measure};
\node[mrisk, below=of p6] (p7) {A maturity level that correlates with nothing observable is a description of documentation, not of capability};
\foreach \a/\b in {p1/p2,p2/p3,p3/p4,p4/p5,p5/p6} \draw[mflow] (\a) -- (\b);
\draw[mflow, densely dotted] (p6) -- (p7);
\coordinate (pl) at ($(p1.west)-(2.6mm,0)$);
\draw[mback] (p3.west) -- (p3.west -| pl)
  -- node[mlbl,rotate=90]{phase orphaned} (pl) -- (p1.west);
\end{tikzpicture}
\caption{Validating a proposed process or maturity model, in the sense
\Cref{sec:operations} argues for. The two gates are what separate a model from a
diagram. Step~2 is the harmonisation discipline the digital-forensics literature arrived at
after accumulating models faster than it could reconcile them; step~4 is the layer below,
where the process is only as sound as the tools it delegates to. Step~6 is the one this
review found most often missing across operational research generally.}
\label{fig:forensic}
\end{figure}
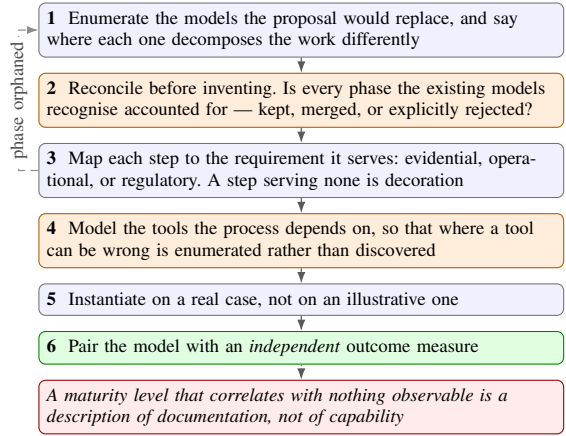

The protocol is illustrated in \Cref{fig:forensic}.

\section{Cross-Cutting Concerns}
\label{sec:crosscutting}

\subsection{Validity, stated as four questions}

Every study in every family should answer four questions explicitly.

\begin{itemize}
  \item \textbf{Construct validity.} Does the measure correspond to the concept? A model's
        $F_1$ on a benchmark is a construct measure of ``ability to separate labelled
        classes in that benchmark'', which is not the same construct as ``detects intrusions
        in an enterprise''. Naming the gap is the whole of the obligation.
  \item \textbf{Internal validity.} Could the observed effect have another cause? Data
        leakage, ordering effects, an uncontrolled confound between conditions, an expert
        who both supplies parameters and validates outputs.
  \item \textbf{External validity.} To which population does the result extend? State the
        population; do not let the reader infer it from the abstract's phrasing.
  \item \textbf{Conclusion validity.} Does the analysis support the inference? Multiple
        comparisons without correction, single-run differences, and effect sizes reported
        without intervals are the recurring problems.
\end{itemize}

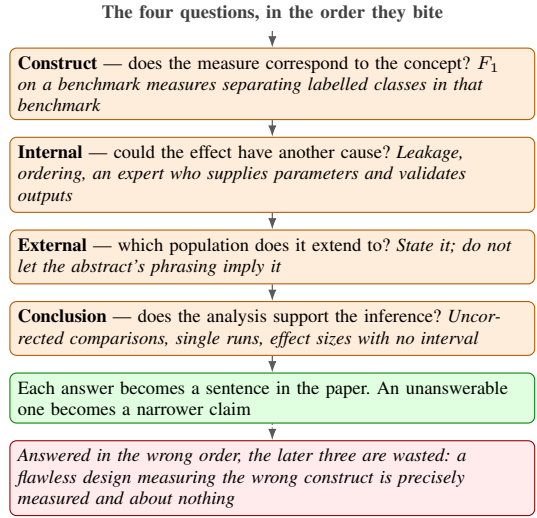
\begin{figure}[t]
\centering
\begin{tikzpicture}[
  node distance=2.2mm,
  mstep/.append style={text width=0.76\columnwidth},
  mgate/.append style={text width=0.76\columnwidth},
  mout/.append style={text width=0.76\columnwidth},
  mrisk/.append style={text width=0.76\columnwidth}]
\node[mhdr] (vh) {The four questions, in the order they bite};
\node[mgate, below=of vh] (v1) {\textbf{Construct} --- does the measure correspond to the concept? \emph{$F_1$ on a benchmark measures separating labelled classes in that benchmark}};
\node[mgate, below=of v1] (v2) {\textbf{Internal} --- could the effect have another cause? \emph{Leakage, ordering, an expert who supplies parameters and validates outputs}};
\node[mgate, below=of v2] (v3) {\textbf{External} --- which population does it extend to? \emph{State it; do not let the abstract's phrasing imply it}};
\node[mgate, below=of v3] (v4) {\textbf{Conclusion} --- does the analysis support the inference? \emph{Uncorrected comparisons, single runs, effect sizes with no interval}};
\node[mout, below=of v4] (v5) {Each answer becomes a sentence in the paper. An unanswerable one becomes a narrower claim};
\node[mrisk, below=of v5] (v6) {Answered in the wrong order, the later three are wasted: a flawless design measuring the wrong construct is precisely measured and about nothing};
\foreach \a/\b in {vh/v1,v1/v2,v2/v3,v3/v4,v4/v5} \draw[mflow] (\a) -- (\b);
\draw[mflow, densely dotted] (v5) -- (v6);
\end{tikzpicture}
\caption{The validity questions every family owes, drawn in order because the order is not
arbitrary. Construct validity is first because it can invalidate the study outright;
conclusion validity is last because it is the only one that can be repaired after the data
are collected. The italic text in each box is the failure this review encountered most often
for that question.}
\label{fig:validity}
\end{figure}

These threats to validity are mapped in \Cref{fig:validity}.

\subsection{Reporting standards, by family}
\label{sec:standards}

\Cref{tab:standards} maps the families onto the reporting instruments that exist for them.
Two features of the table are the reason to print it. First, most of the usable standards
were written outside computer security, in medicine, epidemiology and the social
sciences, and transfer with a change of vocabulary rather than a change of substance;
declining to use one because it says ``participants'' where a security paper says
``hosts'' is not a methodological objection. Second, five of the eleven families have no
reporting standard that we could locate, and those five, threat modeling, simulation, language-model
evaluation, testbeds, and design science, are, not coincidentally, the families in which
this review found protocol details most often unreported. We did not run a systematic
search for candidate standards in the empty rows, so each records an absence in our
retrieval rather than a demonstrated absence in the literature. Where the row is empty, the
family's checklist in the corresponding section is the substitute, and it is a weaker one:
a checklist an author may not have read cannot do what a standard a reviewer expects does.

\begin{table}[t]
\caption{Reporting standards by family. Empty rows record standards we could not locate, not a searched-out absence.}
\label{tab:standards}
\centering
\footnotesize
\renewcommand{\arraystretch}{1.15}
% Ragged right in both columns: justified p-columns stretch a short entry such
% as the CHERRIES row across the full measure, and hyphenate the last row's
% label ("ev-idence") to fill a line it did not need to fill.
\begin{tabular}{@{}L{0.26\columnwidth}L{0.62\columnwidth}@{}}
\hline
\textbf{Family} & \textbf{Instrument} \\
\hline
A Secondary & PRISMA 2020~\cite{page2021prisma}; PRISMA-P for the protocol~\cite{moher2015preferred}; PRISMA-ScR for scoping reviews~\cite{tricco2018prisma} \\
B Design science & --- (use the evaluation ladder, \Cref{fig:dsrladder}) \\
C Qualitative & COREQ for interviews and focus groups~\cite{tong2007consolidated}; SRQR more generally~\cite{obrien2014standards} \\
D Survey & CHERRIES for web-administered instruments~\cite{eysenbach2004improving} \\
E Threat modeling & --- \\
F Simulation & --- \\
G Machine learning & TRIPOD, translated from prediction modelling~\cite{collins2015transparent} \\
H Language models & --- \\
I Measurement & STROBE~\cite{vonelm2007strengthening} \\
J Testbeds & --- \\
K Field experiments & CONSORT~\cite{schulz2010consort} \\
\hline
Rating a body of evidence & GRADE~\cite{guyatt2008grade} \\
\hline
\end{tabular}
\end{table}

\subsection{Reproducibility and artifact sharing}
\label{sec:reproducibility}

Reproducibility in applied security has been measured, not merely lamented. A
reproducibility study of security research assessed how far published work could actually
be re-run~\cite{olszewski2023get}, and an eleven-year review of artifacts at applied
security conferences documents the trajectory of artifact availability and
evaluation~\cite{olszewski2025reproducibility}. Data sharing and reuse practices in the
community have been examined directly, including what researchers do with data they
obtain and what prevents them from releasing it~\cite{crowder2025data}. The practical
minimum for a new study: release code with a pinned dependency specification; release the
exact data split, or the script that regenerates it from a stated source; record
environment and seeds; and separate the artifact that reproduces the paper's figures from
the general-purpose library.

The surrounding practice has a name and a literature. Mendez et al.\ set out open science
for software engineering as a set of concrete, separable commitments, open access, open
data, open source, and pre-registration, together with the objections researchers actually
raise against each and the conditions under which each is achievable~\cite{mendez2020open}.
Treating them as separable matters in security, because the commitments are not equally
available: a study on production telemetry may be unable to open its data and perfectly able
to open its code and pre-register its analysis. Declaring which commitments were met and why
the others were not is more informative than an artifact badge.

Releasing an artifact also creates an obligation that artifact evaluation does not currently
cover. Rani and Rossow examined 509 artifacts from top-tier security venues and report that
many contain insecure code patterns; after static analysis and manual filtering of false
positives, they judge that 41.60\% of the prevalent findings pose potential security concerns
under practical usage, and they propose both a taxonomy for context-aware assessment and a
framework, SAFE, for triaging tool-reported findings~\cite{rani2026security}. The point for
an author is narrow and actionable: artifact evaluation asks whether the artifact
\emph{works}, not whether it is safe for a stranger to run, and those are different
questions. Anything shipped with a research paper should at minimum have its dependencies
pinned, its network behaviour documented, and any credential, key or permissive default
removed before release.

\subsection{Disclosure as part of the method}

Where a study finds a live weakness, disclosure is not an epilogue to the research; it is a
step in it, with its own timeline and its own failure modes. Ayala et al.\ study the
mechanism empirically, analysing 3{,}798 reviewed GitHub security advisories and 4{,}033
disclosed open-source bug-bounty reports to trace how a vulnerability propagates from
reporter to maintainer to dependent project to the global vulnerability
databases~\cite{ayala2025investigating}. What they find in that pipeline is where the delay
lives: missing or late CVE assignment means dependent projects are not notified when they
need to be. A researcher who plans to disclose should design for that pipeline rather than
assume it, identify who can actually issue an identifier for the affected component, start
the clock at report rather than at publication, and state the disclosure timeline in the
paper so that a reader can judge whether publication preceded the fix.

\subsection{Ethics and legal constraints}
\label{sec:ethics}

Security research touches systems and people who did not consent to being studied. The
Menlo Report's application of ethical principles to information and communication
technology research, respect for persons, beneficence, justice, respect for law and
public interest, is the field's reference
framework~\cite{dittrich2013applying}. In enterprise research specifically, three
recurring issues need addressing in the paper itself: employee deception in phishing
studies, which requires approval and debriefing; access to production telemetry, which
requires a data-handling agreement and usually irreversible de-identification; and
disclosure, where a study that finds a live weakness incurs an obligation that precedes
publication.

The Menlo principles tell you which considerations are in play; they do not adjudicate
between them when they conflict, and in security they routinely do. Kohno et al.\ address
exactly that gap, introducing ethical frameworks, consequentialist, deontological and
others, to a security audience and working through hypothetical dilemmas designed so that
different frameworks give different answers~\cite{kohno2023ethical}. The reason to read it
is not to adopt a framework but to be able to say which one a decision rests on. ``We
obtained approval'' records that a committee agreed; it does not record the reasoning, and
a paper whose ethical argument is a procedural fact has not made an argument. Where a study
imposed a cost on someone, deceived employees, degraded a service, collected data about
non-participants, name the cost, name the benefit set against it, and name the principle
under which the trade was made.

\subsection{From result to claim}

The final cross-cutting discipline is the one most often skipped. Every claim in a paper
should trace to a specific artifact, a table, a run, a transcript, a log, and the
strength of the wording should match the strength of that trace. In practice this means
maintaining a claim-to-evidence table alongside the draft, in which each claim names its
supporting artifact and the verification status of that artifact, and reserving
strong comparative language for claims whose supporting evidence has been checked. Where
the artifact is missing or ambiguous, the correct action is to weaken the claim or to ask
the party who holds the data, not to write around the gap. This is the discipline that
would have caught the unreconciled screening figures described in \Cref{sec:seed-audit}.

\section{Reading the Contradictions}
\label{sec:contradiction}

The corpus contains one contradiction sharp enough to be instructive, and working through
it teaches more about evaluation design than any checklist.

\subsection{The contradiction}

Four careful comparative studies rank intrusion-detection algorithms differently.

\begin{itemize}
  \item A comparative study of machine-learning intrusion-detection methods found ensemble
        methods generally strong, while naive Bayes showed lower accuracy on learned data
        but trained faster and had advantages in recognising new attack
        types~\cite{zhang2022comparative}.
  \item A controlled comparison of four deep-learning architectures across KDD~99, NSL-KDD,
        CIC-IDS2017 and CIC-IDS2018 found deep feed-forward networks best on accuracy,
        $F_1$, training time and inference time, with autoencoders and deep belief networks
        not exceeding supervised feed-forward
        networks~\cite{gamage2020deep}.
  \item A recurrent-network intrusion-detection system was reported as superior to J48,
        artificial neural networks, random forest and support vector machines in both
        binary and multiclass classification~\cite{yin2017deep}.
  \item A later comparison found convolutional and recurrent models reaching 98\% accuracy
        while random forest reached 99.9\%~\cite{ali2025deep}.
\end{itemize}

These cannot all be descriptions of a stable ranking.

\subsection{The resolution}

The contradiction is not resolvable by choosing a winner, and attempting to do so is the
error. The results differ because the studies differ in dataset composition, feature
representation, preprocessing, network context, class balance, architecture and
hyperparameter budget, and each of those is known to move the reported number by more
than the margins that separate the studies. What the set of studies establishes jointly is
therefore not a ranking but a constraint: \emph{headline accuracy on a benchmark is not a
stable estimate of enterprise performance}, and a paper that reports one as though it were
has overstated its result regardless of how carefully the experiment was run. The limit
of the argument should be stated with it: attributing the inversion to any single design
factor would require re-running representative models under one harmonised protocol,
which this review does not do (\Cref{sec:gaps}). The claim is that the joint design
variation is sufficient to account for margins of the size observed, not that any one
factor has been isolated.

The hybrid-ensemble case illustrates the limit precisely. Combining XGBoost, random forest,
a graph neural network, long short-term memory and autoencoders under SMOTE and weighted
soft voting, with five-fold cross-validation and an independent benchmark, produced
accuracy, precision, recall and $F_1$ approaching
100\%~\cite{almuhanna2025deep}. The internal comparison is stronger than most. And the
study itself reports computational overhead, inference latency, overfitting risk and the
absence of explicit evaluation against unseen attack types. Stronger validation improved
credibility without removing deployment uncertainty, which is exactly the right outcome to
expect and exactly the outcome that a reader in a hurry will misread.

Publication bias is a plausible additional contributor, positive technical results are
more likely to be written up than failed implementations, though the available data do not
permit its magnitude to be estimated here.

\subsection{What a researcher should do about it}

\begin{enumerate}
  \item Evaluate across multiple datasets, and report per-dataset results rather than an
        average that conceals the variation.
  \item Use a temporally consistent split whenever the claim concerns future
        performance~\cite{pendlebury2018tesseract}.
  \item Report the false-alarm rate at a realistic base rate, not only recall
        (\Cref{eq:bayes}).
  \item Include a simple baseline. Where random forest reaches 99.9\% on a benchmark, that
        benchmark is telling you something about itself.
  \item Report computational cost alongside accuracy, since the deployment decision trades
        them off.
  \item Test against attack types absent from training, and report the degradation.
\end{enumerate}

\section{Three Worked Protocols}
\label{sec:recipes}

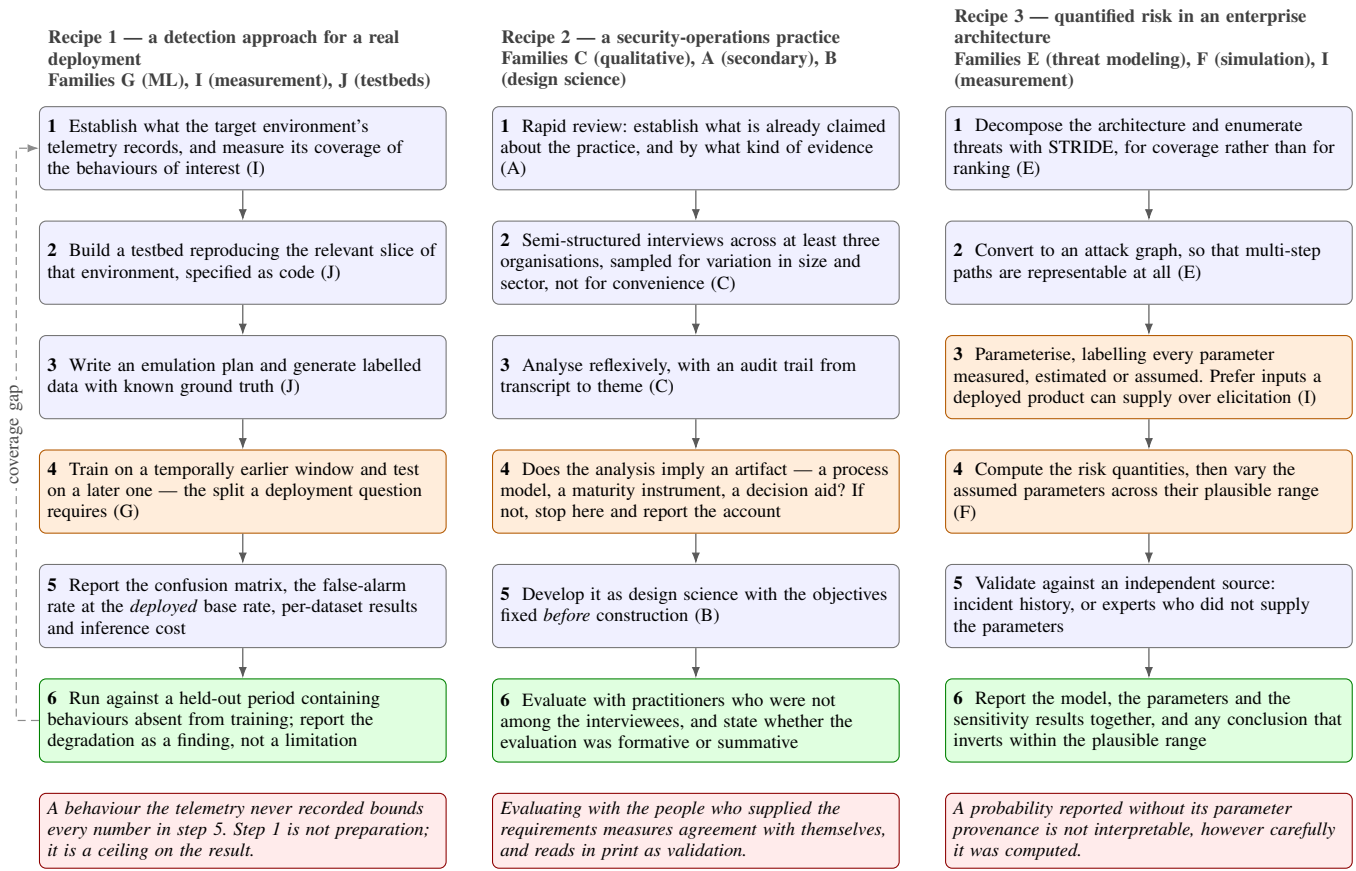
\begin{figure*}[t]
\centering
\begin{tikzpicture}[
  mstep/.append style={text width=0.285\textwidth, minimum height=11mm},
  mgate/.append style={text width=0.285\textwidth, minimum height=11mm},
  mout/.append style={text width=0.285\textwidth, minimum height=11mm},
  mrisk/.append style={text width=0.285\textwidth, minimum height=9mm}]

% --- Recipe 1: detection for a real deployment (G, I, J) ---
\node[mstep] (r1a) {\sn{1} Establish what the target environment's telemetry records, and measure its coverage of the behaviours of interest (I)};
\node[mstep, below=4mm of r1a] (r1b) {\sn{2} Build a testbed reproducing the relevant slice of that environment, specified as code (J)};
\node[mstep, below=4mm of r1b] (r1c) {\sn{3} Write an emulation plan and generate labelled data with known ground truth (J)};
\node[mgate, below=4mm of r1c] (r1d) {\sn{4} Train on a temporally earlier window and test on a later one --- the split a deployment question requires (G)};
\node[mstep, below=4mm of r1d] (r1e) {\sn{5} Report the confusion matrix, the false-alarm rate at the \emph{deployed} base rate, per-dataset results and inference cost};
\node[mout, below=4mm of r1e] (r1f) {\sn{6} Run against a held-out period containing behaviours absent from training; report the degradation as a finding, not a limitation};
\node[mrisk, below=4mm of r1f] (r1r) {A behaviour the telemetry never recorded bounds every number in step~5. Step~1 is not preparation; it is a ceiling on the result.};

% --- Recipe 2: a security-operations practice (C, A, B) ---
\node[mstep, anchor=north west] (r2a) at ($(r1a.north east)+(6mm,0)$) {\sn{1} Rapid review: establish what is already claimed about the practice, and by what kind of evidence (A)};
\node[mstep, below=4mm of r2a] (r2b) {\sn{2} Semi-structured interviews across at least three organisations, sampled for variation in size and sector, not for convenience (C)};
\node[mstep, below=4mm of r2b] (r2c) {\sn{3} Analyse reflexively, with an audit trail from transcript to theme (C)};
\node[mgate, below=4mm of r2c] (r2d) {\sn{4} Does the analysis imply an artifact --- a process model, a maturity instrument, a decision aid? If not, stop here and report the account};
\node[mstep, below=4mm of r2d] (r2e) {\sn{5} Develop it as design science with the objectives fixed \emph{before} construction (B)};
\node[mout, below=4mm of r2e] (r2f) {\sn{6} Evaluate with practitioners who were not among the interviewees, and state whether the evaluation was formative or summative};
\node[mrisk, below=4mm of r2f] (r2r) {Evaluating with the people who supplied the requirements measures agreement with themselves, and reads in print as validation.};

% --- Recipe 3: quantified risk in an architecture (E, F, I) ---
\node[mstep, anchor=north west] (r3a) at ($(r2a.north east)+(6mm,0)$) {\sn{1} Decompose the architecture and enumerate threats with STRIDE, for coverage rather than for ranking (E)};
\node[mstep, below=4mm of r3a] (r3b) {\sn{2} Convert to an attack graph, so that multi-step paths are representable at all (E)};
\node[mgate, below=4mm of r3b] (r3c) {\sn{3} Parameterise, labelling every parameter measured, estimated or assumed. Prefer inputs a deployed product can supply over elicitation (I)};
\node[mgate, below=4mm of r3c] (r3d) {\sn{4} Compute the risk quantities, then vary the assumed parameters across their plausible range (F)};
\node[mstep, below=4mm of r3d] (r3e) {\sn{5} Validate against an independent source: incident history, or experts who did not supply the parameters};
\node[mout, below=4mm of r3e] (r3f) {\sn{6} Report the model, the parameters and the sensitivity results together, and any conclusion that inverts within the plausible range};
\node[mrisk, below=4mm of r3f] (r3r) {A probability reported without its parameter provenance is not interpretable, however carefully it was computed.};

% column headers
\node[mhdr, text width=0.285\textwidth, align=left, anchor=south west] at ($(r1a.north west)+(0,1mm)$)
  {Recipe 1 --- a detection approach for a real deployment\\Families G (ML), I (measurement), J (testbeds)};
\node[mhdr, text width=0.285\textwidth, align=left, anchor=south west] at ($(r2a.north west)+(0,1mm)$)
  {Recipe 2 --- a security-operations practice\\Families C (qualitative), A (secondary), B (design science)};
\node[mhdr, text width=0.285\textwidth, align=left, anchor=south west] at ($(r3a.north west)+(0,1mm)$)
  {Recipe 3 --- quantified risk in an enterprise architecture\\Families E (threat modeling), F (simulation), I (measurement)};

\foreach \a/\b in {r1a/r1b,r1b/r1c,r1c/r1d,r1d/r1e,r1e/r1f,
                   r2a/r2b,r2b/r2c,r2c/r2d,r2d/r2e,r2e/r2f,
                   r3a/r3b,r3b/r3c,r3c/r3d,r3d/r3e,r3e/r3f}
  \draw[mflow] (\a) -- (\b);

% Discovering in step 6 that the telemetry cannot see the behaviour returns the
% study to step 1, not to the model.
\coordinate (rl) at ($(r1a.west)-(3mm,0)$);
\draw[mback] (r1f.west) -- (r1f.west -| rl)
  -- node[mlbl,rotate=90]{coverage gap} (rl) -- (r1a.west);
\end{tikzpicture}
\caption{The three worked protocols, side by side. Read down a column for one study; read
across for what the three have in common. Each begins by fixing what can be observed or
claimed at all (steps~1--2), forces a decision that determines what the result will license
(the orange gate), and ends with an output whose required contents are stated rather than
left to the author (green). The red band is the recipe's characteristic failure: in each case
it is a defect introduced early that no later rigour repairs, which is why the ordering is
part of the protocol.}
\label{fig:recipes}
\end{figure*}

The families combine. Three templates cover a large share of realistic enterprise studies.
\Cref{fig:recipes} places them side by side, in the shared notation used for the individual
families, so that the composition is legible as a composition: each recipe is a path through
three families rather than a fourth methodology.

\subsection{Recipe 1: Evaluating a detection approach for a real deployment}

\emph{Families G, I, J.} Establish the telemetry available in the target environment and
measure its coverage of the behaviours of interest (\S\ref{sec:measurement}). Build a
testbed that reproduces the relevant portion of the environment and specify it as code
(\S\ref{sec:infrastructure}). Write an emulation plan and generate labelled data with known
ground truth. Train on a temporally earlier window and test on a later one
(\S\ref{sec:temporal}). Report the confusion matrix, the false-alarm rate at the
environment's actual base rate, per-dataset results, and inference cost. Then run the
model against a held-out period containing behaviours absent from training, and report the
degradation as a finding rather than a limitation.

\subsection{Recipe 2: Studying a security-operations practice}

\emph{Families C, A, B.} Begin with a rapid review to establish what is already claimed
(\S\ref{sec:secondary}). Conduct semi-structured interviews with practitioners across at
least three organisations, sampling for variation in size and sector rather than for
convenience (\S\ref{sec:qualitative}). Analyse reflexively, with an audit trail. If the
analysis suggests an artifact, a process model, a maturity instrument, a decision
aid, develop it as design science with objectives fixed in advance
(\S\ref{sec:dsr}), and evaluate it with practitioners who were not among the
interviewees. Report the formative-summative distinction honestly.

\subsection{Recipe 3: Quantifying risk in an enterprise architecture}

\emph{Families E, F, I.} Decompose the architecture and enumerate threats with STRIDE to
establish coverage (\S\ref{sec:threat}). Convert to an attack graph so that multi-step
paths are representable, and parameterise it, labelling each parameter as measured,
estimated or assumed. Where deployed security products can supply inputs automatically,
prefer that to elicitation~\cite{sato2025malcoda}. Compute the risk quantities and then run
sensitivity analysis across the assumed parameters; report any conclusion that inverts
within their plausible range. Validate against an independent source, incident history, or
the judgement of experts who did not supply the parameters. Report the model, the
parameters and the sensitivity results together; a probability without its parameter
provenance is not interpretable.

\section{Gaps and a Research Agenda}
\label{sec:gaps}

\subsection{Gaps in the field}

\begin{enumerate}
  \item \textbf{Evaluation in continuously operating enterprises.} The largest gap.
        Detection studies rest on KDD~99, NSL-KDD, CIC-IDS2017, CIC-IDS2018, synthetic
        attacks and simulated traffic; risk and cyber-physical studies rest on modelled
        systems. Work that combines benchmark reproducibility with longitudinal enterprise
        telemetry, multiple independent organisations, changing attack distributions and
        explicit unseen-attack testing would resolve more open questions than any further
        architecture comparison.
  \item \textbf{Harmonised comparison protocols.} The contradiction in
        \Cref{sec:contradiction} cannot be resolved without shared preprocessing, feature
        selection, imbalance treatment, hyperparameter reporting, cost measurement and
        train--test protocol.
  \item \textbf{Validated outcome measures for operations.} Incident-response and awareness
        research lacks measures that link exercises, training and intelligence integration
        to response performance. Until those exist, maturity comparisons across
        organisations are not interpretable.
  \item \textbf{Interaction-level threat modeling.} STRIDE's documented blind spot, threats
        arising from component interaction rather than from components, remains
        open~\cite{khan2017stridebased}.
  \item \textbf{Uncertainty reporting in probabilistic risk models.} Models should report
        their data requirements and their sensitivity to incomplete telemetry, which is the
        normal operating condition.
  \item \textbf{Under-represented contexts.} Small and medium-sized enterprises,
        non-technical employees, small security-operations teams, and enterprises outside
        finance, healthcare, critical infrastructure and
        transport~\cite{chaudhary2023quest,vanderkleij2017computer}. A ransomware-mitigation
        model developed and validated for micro and small enterprises is one of the few
        direct engagements with this population~\cite{biggi2025development}.
  \item \textbf{Evaluation methodology for agentic systems}, which is younger than the
        systems it is asked to assess (\S\ref{sec:llm}).
\end{enumerate}

\subsection{Gaps in this review}

We under-covered three facets and name them so that a reader does not mistake absence for
evidence of absence: operational security metrics such as mean time to detect and mean time
to respond; insider-threat study design; and the design of studies that recruit adversaries
or observe them at scale. Deception measurement and forensic process validation are treated
here only as far as their study designs go (\S\ref{sec:measurement},
\S\ref{sec:operations}); neither receives the family-level treatment given to the eleven,
and both would support one. The same is true of the three facets above: each merits the
treatment given to the eleven families here.

\section{Conclusion}
\label{sec:conclusion}

Enterprise cybersecurity research is best conducted through a combination of methodologies
rather than a single all-purpose design. Secondary research organises fragmented knowledge;
design science produces and tests artifacts; controlled experiments compare detection
methods; threat modeling and probabilistic analysis structure risk; simulation permits safe
evaluation of complex and cyber-physical systems; qualitative, mixed and longitudinal
approaches reveal the organisational, cognitive and behavioural conditions that technical
metrics do not reach.

The technical evidence is promising and uneven. Reported detection performance is high but
the ranking among approaches is inconsistent across studies; we argue the inconsistency
is better explained by evaluation design than by the algorithms, since the studies
differ jointly --- in dataset, preprocessing, class balance and tuning budget ---
by more than the margins that separate them, though no single factor has been
isolated (\Cref{sec:contradiction}). Risk-modeling approaches produce usable probabilities in
virtual enterprise networks and await broader replication~\cite{sato2025malcoda}.
Organisational studies supply the complementary evidence that communication, decision
load, human factors, intelligence integration and temporal variation in behaviour shape
security outcomes~\cite{ahmad2015case,cram2024time,vanderkleij2022developing}.

The unresolved question is whether these methods retain their reported effectiveness under
changing enterprise conditions, unseen attacks, incomplete telemetry, shifting employee
behaviour, legacy systems and resource constraints. Answering it requires triangulating
models, operational data, practitioner evidence and longitudinal evaluation. Two habits do
most of the work in getting there, and neither requires new technique: state the population
a result generalises to, and state the observation that would have shown the approach did
not work. A literature that did both consistently would resolve most of the contradictions
catalogued here without running a single additional experiment.

\appendices

\section{Consolidated Reporting Checklists}
\label{app:checklists}

\Cref{tab:checklist} consolidates the per-family checklists into the items most frequently
absent from published work in this corpus. It is intended for use both when writing and
when reviewing.

\begin{table*}[t]
\caption{Consolidated reporting items, by family. Items are those most often absent in the reviewed corpus.}
\label{tab:checklist}
\centering
\footnotesize
\begin{tabularx}{\textwidth}{@{}l Y Y@{}}
\toprule
\textbf{Family} & \textbf{Report always} & \textbf{The item most often missing} \\
\midrule
Secondary research & Protocol, queries with dates, databases, snowballing, dual screening, PRISMA flow & Full-text exclusion reasons; a flow that reconciles \\
Design science & Measurable objectives fixed before build; artifact specification; evaluation type & Whether evaluation was formative or summative; the incumbent baseline \\
Qualitative & Sampling rationale, interview guide, coding procedure, reflexivity & Consistency between the design's rigour criterion and the statistic reported \\
Survey & Population versus frame, validated scales, power analysis, measurement model & The population the estimate generalises to \\
Threat modeling & Decomposition, method scope limit, parameter sources & Sensitivity analysis; separation of parameter suppliers from validators \\
Simulation & Model specification, parameter classification, replications & Verification distinct from validation; sensitivity on assumed parameters \\
Machine learning & Data provenance, split type, preprocessing placement, confusion matrix, cost & False-alarm rate at a realistic base rate; temporal consistency; seed variance \\
Language-model and agentic & Model version and date, attempts per task, judge calibration, environment & Contamination control; cost per task \\
Measurement & Ground-truth source, exact configuration, coverage before efficacy & Separation of logged, detected and alerted \\
Infrastructure & Environment as code, fidelity claim, containment, emulation plan & Which properties of reality the testbed does \emph{not} reproduce \\
Field experiment & Randomisation unit, pre-specified outcome, intention-to-treat, ethics & Contamination between conditions; follow-up interval \\
\bottomrule
\end{tabularx}
\end{table*}

\bibliographystyle{IEEEtran}
\bibliography{references}

\end{document}